\documentclass[preprint,11pt]{aastex}

\shorttitle{Disk of L1527}
\shortauthors{Tobin et al.}
\newcommand{\mum}{\mbox{$\mu$m}}
\newcommand{\nthp}{\mbox{N$_2$H$^+$}}
\newcommand{\ntdp}{\mbox{N$_2$D$^+$}}

\newcommand{\hcop}{\mbox{HCO$^+$}}

\newcommand{\kmspc}{\mbox{km s$^{-1}$ pc$^{-1}$ }}
\newcommand{\kms}{\mbox{km s$^{-1}$}}

\begin{document}

\title{Modeling the Resolved Disk Around the Class 0 Protostar L1527}
\author{John J. Tobin\altaffilmark{1,8},  Lee Hartmann\altaffilmark{2}, Hsin-Fang Chiang\altaffilmark{3,4}, David J. Wilner\altaffilmark{5},
Leslie W. Looney\altaffilmark{3}, Laurent Loinard\altaffilmark{6,7}, Nuria Calvet\altaffilmark{2}, Paola D'Alessio\altaffilmark{6}}

\altaffiltext{1}{National Radio Astronomy Observatory, Charlottesville, VA 22903; jtobin@nrao.edu}
\altaffiltext{2}{Department of Astronomy, University of Michigan, Ann Arbor, MI 48109}
\altaffiltext{3}{Department of Astronomy, University of Illinois, Urbana, IL 61801 }
\altaffiltext{4}{Institute for Astronomy and NASA Astrobiology Institute, University
of Hawaii at Manoa, Hilo, HI 96720}
\altaffiltext{5}{Harvard-Smithsonian Center for Astrophysics, Cambridge, MA 02138}
\altaffiltext{6}{Centro de Radioastronom{\'\i}a y Astrof{\'\i}sica, UNAM, Apartado Postal
3-72 (Xangari), 58089 Morelia, Michoac\'an, M\'exico}
\altaffiltext{7}{Max-Planck-Institut f\"ur Radioastronomie, Auf dem H\"ugel 69, 53121 Bonn, Germany}

\altaffiltext{8}{Hubble Fellow}
\begin{abstract}

We present high-resolution sub/millimeter interferometric 
imaging of the Class 0 protostar L1527 IRS (IRAS 04368+2557) at $\lambda$ = 870 \micron\ and
3.4 mm from the Submillimeter Array (SMA) and Combined Array for Research
in Millimeter Astronomy (CARMA). We detect the signature of an edge-on disk
surrounding the protostar with an observed diameter of 180 AU in the sub/millimeter images.
The mass of the disk is estimated to be 0.007 $M_{\sun}$, assuming optically thin, isothermal
dust emission. The millimeter spectral index is observed
to be quite shallow at all the spatial scales probed; $\alpha$ $\sim$ 2, implying 
a dust opacity spectral index $\beta$ $\sim$ 0.
We model the emission from the disk and surrounding envelope using Monte Carlo
radiative transfer codes, simultaneously fitting the sub/millimeter visibility amplitudes,
sub/millimeter images, resolved L\arcmin\ image, spectral energy distribution, and mid-infrared 
spectrum. The best fitting model has a disk radius of R = 125 AU, is
highly flared ($H$ $\propto$ $R^{1.3}$), has a radial density profile 
$\rho$ $\propto$  $R^{-2.5}$, and has a mass of 0.0075 $M_{\sun}$.
The scale height at 100 AU is 48 AU, about a factor of two 
greater than vertical hydrostatic equilibrium.
The resolved millimeter observations indicate that disks may
grow rapidly throughout the Class 0 phase.
The mass and radius of the young disk around L1527 is comparable 
to disks around pre-main sequence stars; however, the disk is considerably more 
vertically extended, possibly due to a combination of lower protostellar mass, infall onto
the disk upper layers, and little settling of $\sim$1 \micron-sized dust grains.

\end{abstract}


\section{Introduction}

In the earliest stages of star formation, the Class 0/I phases \citep{andre1993,lada1987},
a newborn protostar is embedded within a dense infalling envelope of gas and dust. 
Disks are thought to naturally form in these systems due to conservation of angular 
momentum in the collapsing cloud; several analytic models have been developed
that explain the formation of disks within infalling envelopes \citep{cassen1981, tsc1984}. 
Observational characterization of disks around Class 0/I objects has been difficult because
emission from the surrounding envelope is entangled with that of the disk \citep[e.g.][]{chiang2008,jorgensen2009}. 
However, the disks around more evolved young stars without envelopes, Class II
sources, have been studied in great detail in recent
years, deriving their surface density distributions, radii, masses, and
dust opacity spectral indicies from sub/millimeter interferometric imaging 
\citep[e.g.][]{dutrey1996, kitamura2002,andrews2009, isella2009,ricci2010,kwon2011}. 
Gaps and holes have also been discovered in disks around pre-main sequence 
stars \citep{calvet2005,pietu2006,espaillat2007,espaillat2008, hughes2009}, which are interpreted
as signs of planets formation, suggesting that disks are ultimately
cleared by the accretion of material onto proto-planets. 

A relatively clear picture of pre-main sequence disk properties and how disks 
are dispersed has been gained; however,
we still do not have a clear observational picture of how the life of a disk begins.
Large disks have been clearly shown to exist in the Class I phase 
\citep{eisner2012,wolf2008,padgett1999,launhardt2001,jorgensen2009,takakuwa2012,eisner2012},
but firm detections of Class 0 protostellar disks have 
remained elusive \citep{looney2003,harvey2003,jorgensen2009,chiang2008,chiang2012}
and a controversy exists as to their expected properties. Specifically, it has remained uncertain 
whether or not large disks form during this phase and their expected masses. 

Hydrodynamic simulations show that disks can grow to large radii rather quickly in the absence of strong magnetic
fields \citep{yorke1999} and are are often massive enough to fragment \citep{vorobyov2010}.
On the other hand, Models of protostellar collapse in magnetized envelopes have difficulty forming
rotationally supported disks under the assumption of ideal magneto-hydrodynamics (MHD)
due to strong magnetic braking as field lines are dragged
inward, this is the so-called "magnetic braking catastrophe" \citep[e.g.][]{allen2003,galli2006,hennebelle2008,mellon2008}.
Recently models that consider non-ideal MHD have been able to form 
small, rotationally-supported disks as the magnetic flux is 
dissipated via Ambipolar Diffusion and Ohmic dissipation \citep{dapp2010,machida2010}.
Disks in these simulations are expected to have radii $<$ 100 AU until the end 
of the Class I phase; however, the robustness of Ohmic dissipation for the
formation of rotationally-supported disks has been questioned \citep{li2011}. Nonetheless,
simulations considering non-idealized initial conditions (i.e. magnetic field mis-alignment with rotation
axis and turbulent cores) are able to form rotationally supported disks under the assumption of ideal MHD \citep{joos2012,seifried2012}.

Despite the large body of theoretical and numerical work toward understanding
disk formation,  only a few Class 0/I 
systems have been observed with sufficient resolution and sensitivity
in the millimeter spectral range to even have the possibility of characterizing the 
properties of disks in this early phase of evolution. A modest resolution 
protostellar survey with the Sub-millimeter Array (SMA) by \citet{jorgensen2009}
found that the compact dust continuum emission from protostars is 
consistent with the presence of disks having 
masses between 0.002 and 0.5 $M_{\sun}$. However, modeling of millimeter continuum data from Class 0
protostars indicates that non-uniform density structure (i.e. radial density enhancements at small-scales)
can reproduce the observed data without a disk \citep{chiang2008}.
\citet{maury2010} carried out a high-resolution study toward five protostars in Taurus and Perseus
with the Plateau de Bure Interferometer (PdBI) and did not detect disks or close binaries.
The authors then argued that the formation of large disks and fragmentation are 
suppressed by magnetic fields during collapse.

Contrary to theoretical expectations, a clear 
detection of a Class 0 disk is found toward the protostar L1527 in Taurus 
(d=140 pc) \citep[][hereafter Papers I \& II]{tobin2010b,tobin2012}.
Paper I presented high resolution L$^{\prime}$-band (3.8$\mu$m) observations of L1527 and found the 
apparent signature of an edge-on protostellar disk in near-infrared scattered light.
Follow-up observations with the SMA and Combined Array for Research in Millimeter-wave Astronomy (CARMA)
at $\sim$0\farcs3 resolution found the dust emission to be resolved and elongated in the same direction
as the near-infrared dark lane \citep{tobin2012}. Moreover, $\sim$1\arcsec\ resolution
observations of the $^{13}$CO detected the signature of Keplerian rotation in the disk via 
spectro-astrometry, enabling the protostellar mass to be constrained to be 0.19$\pm$0.04 $M_{\sun}$.

In this paper, we present radiative transfer modeling of the disk around L1527. We attempt to 
construct a realistic model of the protostellar disk embedded within its envelope by simultaneously
considering  the disk continuum emission, multi-wavelength SED, \textit{Spitzer} IRS spectrum, and L\arcmin\ image.
We find that the data are best reproduced by a highly-flared R = 125 AU, M = 0.0075 $M_{\sun}$ disk, with a very shallow
dust emissivity spectral index.
Section 2 describes our observations and data reduction, our observational 
results are presented in Section 3, we discuss the modeling in Section 4, 
the results are discussed in Section 5, and conclusions are given in Section 6.

\section{Observations and Data Reduction}

The observations we present are among the highest resolution taken toward a 
Class 0 protostar using CARMA \citep{woody2004} at $\lambda$ = 3.4 mm and
the SMA \citep{ho2004} at $\lambda$ = 870 \micron. The data were taken in multiple configurations 
at both facilities, gaining sensitivity to both large and small-scale structures 
in the protostellar envelope and disk. 

\subsection{CARMA Observations and Data Reduction}

The CARMA 3.4 mm data were taken in all five CARMA configurations between 2008 and 
2010. We discuss the A and B array data
separately from the C, D, and E-array observations due to the different 
calibration techniques. Details of each observation are given in Table 1.

\subsubsection{3.4 mm A \& B-array}
The A-array observations of L1527 were taken on 2010 December 02 during stable conditions with $\sim$4mm of precipitable 
water vapor (pwv). The local oscillator was tuned to $\nu$=90.7510 GHz and the correlator was configured for 
4-bit sampling with 8 - 500 MHz moveable spectral windows, measuring continuum emission. The IF range spans 1 to 9 GHz and 
we arranged windows such that they occupied the 1 to 5 GHz range of IF bandwidth. 
The receivers operate in dual-sideband (DSB) mode and the full instantaneous bandwidth was 8 GHz 
in one polarization. The primary gain calibrator was 3C111,
12.8 degrees away on the sky; 0431+206 was observed as a test source in conjunction with L1527. 3C84 was observed
as the bandpass calibrator, and Neptune was observed as the absolute flux calibrator. The observations were conducted
in a standard loop, observing 3C111 for 3 minutes, L1527 for 7 minutes, 0431+206 for 1 minute, and then repeating.
Pointing was updated periodically using optical pointing correction and radio pointing was done once during the
track.

The B-array observations were taken on 2010 January 02 and 06. The local oscillator was tuned to 
$\nu$=87.27 GHz during both observations and one 500 MHz spectral window was configured for continuum observation.
This yielded 1 GHz of total continuum bandwidth in DSB, with the windows centered on $\nu$=85.5 and 89.05 GHz.
These observations were taken using the old correlator with less overall bandwidth.
The primary gain calibrator was 3C111, 3C123 was observed as test source, 3C84 was the bandpass calibrator,
and Uranus was observed as the flux calibrator in the first track. No flux calibrator was observed during the
second track, so the flux of 3C111 was carried over from the first track because they were observed only 4 days apart.
The observations were conducted in a loop, observing 3C111 for 3 minutes, L1527 for
9 minutes, and 3C123 for 3 minutes.

The CARMA paired antenna calibration system (C-PACS) was used for enhanced atmospheric phase correction
in both the A and B-array observations. Briefly, the C-PACS calibration works
as follows, the eight 3.5 meter dishes from the SZ array are positioned next to the 10.4 meter and 6 meter antennas
at the longest baselines. While the source(s) is being observed, the 3.5 meter dishes are observing
a nearby calibrator at 30 GHz. The short timescale phase variations during the observation
of the source(s) are corrected from the simultaneous calibrator observations by the
3.5 meter antennas. This correction effectively
reduces the atmospheric decorrelation at the longest baselines, so long as the C-PACS calibrator is 
sufficiently close to the source(s). A more detailed explanation of the C-PACS system is given in \citet{perez2010}.
During the A-array observation of L1527, the C-PACS antennas observed the quasar 0428+329 (7.4\degr\ away)
with a flux density of 1.9 Jy at 30 GHz. For the B-array observations, the C-PACS antennas observed the quasar J0403+260 (8.3\degr\ away) 
and had a flux density of 1.6 Jy.

\subsubsection{3.4 mm C, D, \& E-array Observations }
The C-array observations were conducted on 2009 May 28 and 2009 October 11. The local oscillator was tuned
to $\nu$=87.27 GHz and two 500 MHz bands were configured for continuum observation (2GHz DSB). 
3C111 was observed as the gain calibrator, 3C84 and Uranus were the flux 
calibrators, and 0423-013 and 3C84 were the bandpass calibrators. L1527 
was again observed in D-array on 2009 July 28 and 30, as well as 2008 August 31 with the local oscillator tuned to 
$\nu$=87.27 GHz for the 2009 July observations and $\nu$=91.2 GHz for the 2008 August observations.
In both data sets, two 500 MHz bands were configured for continuum observation (2GHz DSB).
 3C111 was observed as the
gain calibrator, Uranus and Mars were the flux calibrators, and 3C111 and 3C84 were the bandpass calibrators.
Lastly, L1527 was observed in E configuration in 2008 October; these observations were taken
in a three-point mosaic to better recover the large-scale emission from the envelope. The correlator
was configured with one band for 500 MHz continuum, the other two bands were set to observe \nthp\ ($J=1\rightarrow0$) 
and \hcop\ ($J=1\rightarrow0$). The \nthp\ and E-array continuum observations were previously presented in \citet{tobin2011}.
 3C111 was observed as the gain calibrator and 3C84 was observed as both the flux and bandpass calibrator.

\subsubsection{Data Reduction}
The data were reduced using the MIRIAD software package \citep{sault1995}. The visibilities were first corrected for
the updated baseline solutions and transmission line length correction. Then the amplitudes and phases were examined
for each baseline, flagging the source observations between points where the calibrator had bad amplitudes or extremely 
high phase variance. The data were then bandpass corrected using the \textit{mfcal} task. The absolute flux calibration
was calculated using the \textit{bootflux} task to determine the flux of 3C111 relative to the absolute flux calibrator.
We compared the visibility amplitudes of L1527 in ranges of overlapping uv-coverage to estimate our flux calibration uncertainty.
There was excellent agreement between the 3.4 mm datasets and we estimate an absolute flux uncertainty of $\sim$10\%.

For the A and B-array data taken with the C-PACS system, the 30 GHz
data were also inspected for bad amplitudes and phases. One C-PACS antenna was offline during
the A-array track, but all eight were operational for the B-array tracks. The C-PACS data were
bandpass corrected using 3C111 and the data were self-calibrated on a timescale of 4 seconds. The time-dependent 
phase correction to apply to the science data was calculated by the \textit{gpbuddy} task and the
corrections were applied using the \textit{uvcal} task with the \textit{atmcal} option. The science array 
gain calibrations were then calculated for 3C111 with the \textit{mselfcal} task and this solution was applied 
to the observations of L1527. The angular separation of the C-PACS calibrator to L1527 was not ideal but, the
correction did reduce the root-mean squared (RMS) phase scatter on 3C111 from $\sim$60\degr\ to $\sim$40\degr, comparable
effectiveness is assumed on the L1527 observations.

The data were then imaged by first computing the inverse Fourier transform of the
data with the \textit{invert} task, weighting by system temperature and creating the dirty map. The dirty map is then CLEANed
using the \textit{mossdi} task and the image is restored and convolved with the CLEAN beam by the \textit{restor} task. The images
were cleaned down to 1.5$\times$ the RMS. The images of the test sources 0431+206 and 3C123 were also reconstructed
to ensure that these secondary calibrators appear as point sources as a check of the millimeter ``seeing''. The test
source images were point sources, confirming the good conditions.

\subsection{Submillimeter Array Observations}

We observed L1527 with the Submillimeter Array (SMA)
at Mauna Kea on 2011 January 5 and 6. The data were taken in very extended configuration
with 7 antennas operating. The first track had excellent phase coherence but the pwv was $\sim$4.0 mm and
the second track had $\sim$2.0 mm pwv, but worse phase coherence. The local oscillator
was tuned to $\nu$=347.02 GHz and the correlator was used in 4 GHz (DSB) mode.
The correlator was configured for 32 channels per chunk, the coarsest resolution mode
for continuum observations, each chunk is 104 MHz wide and there
are 48 correlator chunks. The IF frequency was chosen to overlap with the compact array dataset taken
by the PROSAC project on 2004 December 17 \citep{jorgensen2007}. 3C111 was observed as the gain
calibrator, 0510+180 was observed as a secondary calibrator, 3C279 was the bandpass
calibrator, and Callisto was the absolute flux calibrator.
The data were taken in the following loop: 3C111 was observed for 2.5 minutes, then L1527 was observed for
8 minutes, then 0510+180 was observed for 2.5 minutes, then L1527 again and finishing with 3C111.

The compact array observations were taken on 2004 December 18 and reported in \citet{jorgensen2007}.
Seven antennas were operating during the track; however, only six antennas produced useable data. The correlator had three chunks 
configured for higher resolution spectral line observations and the rest configured for 
continuum measurement; including the spectral line chunks with more channels, the observations had
2 GHz (DSB) of bandwidth. The gain calibrators were 3C111 and 0510+180, Saturn was used for bandpass calibration,
and the flux calibrator was Uranus. The data were taken observing 3C111 for 5 minutes, L1527 for 15 minutes, and
0510+180 for 5 minutes, then L1527 again and repeating the cycle. Details of the observations are given in Table 2.

The data were reduced using the MIR software package, an IDL-based
software package originally developed for the Owens Valley Radio
Observatory and adapted by the SMA group. The data were first corrected for
the updated baseline solutions and then phases and amplitudes on each
baseline were inspected and uncalibrateable data were flagged.
The system temperature correction was then applied to the data and 
the bandpass correction was calculated, trimming three channels
at the edge of each correlator chunk. After bandpass calibration, 
a first-pass gain calibration was performed on 3C111, 0510+180, and Callisto
to measure the calibrator fluxes. The L1527 and 0510+180 data were then 
calibrated for amplitude and phase using only 3C111. The visibility amplitudes of
the compact and very extended configuration data were consistent 
in the regions of uv overlap, despite the different flux calibrators; we estimate
an absolute flux uncertainty of 10\%. The data were then exported to
MIRIAD format for imaging and were imaged using the same technique as 
outlined in the previous section. 0510+180 was additionally imaged and 
found to be a point source, confirming the excellent submillimeter seeing 
conditions for the very extended observations.

\section{Results}

The observations of L1527 at multiple wavelengths and configurations enable us to perform
a broad observational characterization of the dust continuum emission from the protostar. 
We compare and combine our work with previous observations of L1527 taken with Gemini, PdBI, VLA, and
EVLA to derive a more complete understanding of the disk and envelope.

\subsection{Multi-Configuration Images}

\begin{figure}
\begin{center}
\includegraphics[angle=-90, scale=0.4]{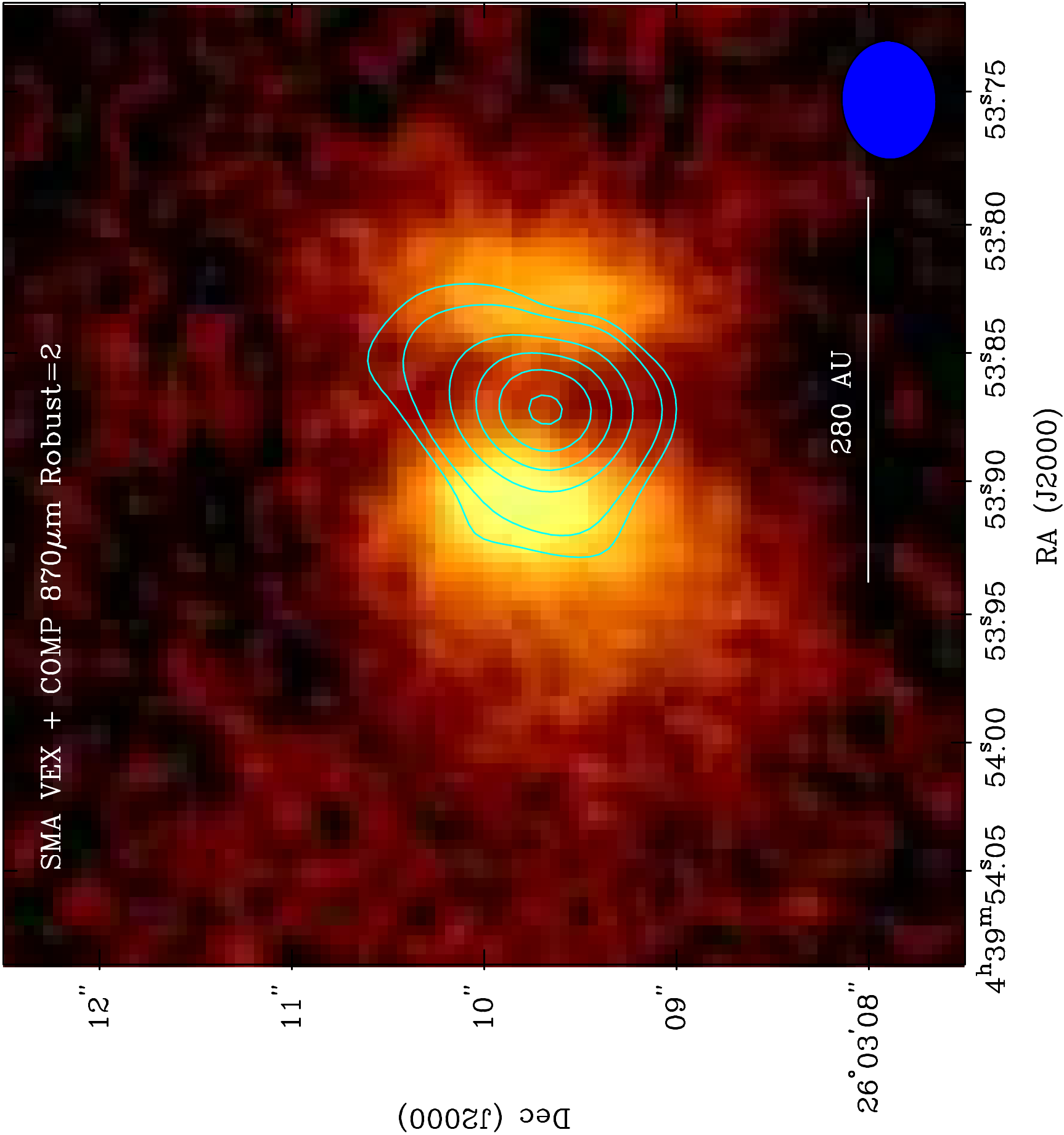}
\includegraphics[angle=-90, scale=0.4]{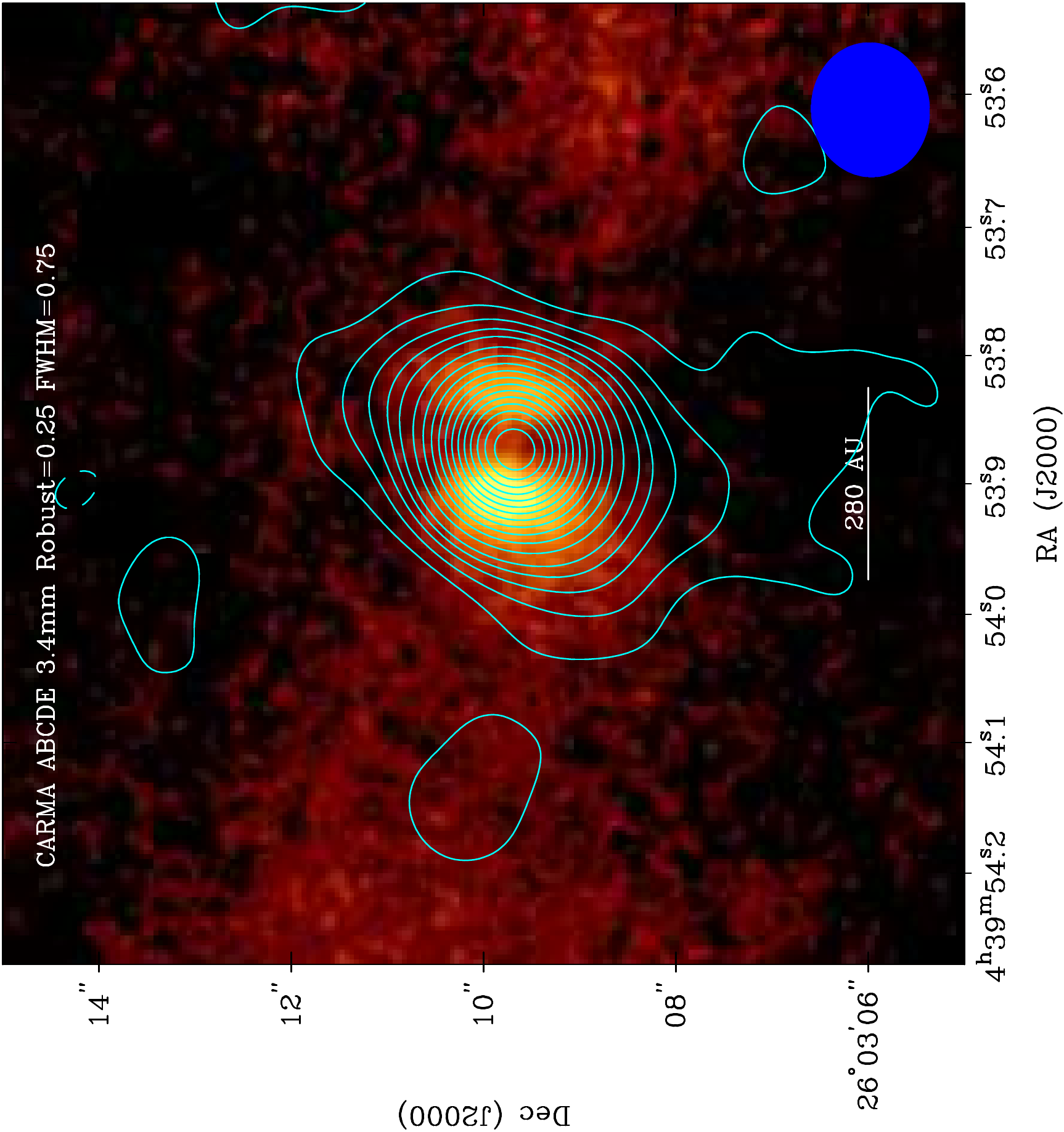}
\end{center}
\caption{Images of L1527 from the SMA at 870 \micron\ (left) and CARMA at 3.4 mm (right) including data from all observed array
configurations. The addition of shorter-spacing data enable larger scale emission to be recovered. The larger-scale
emission is more extended in the direction of the dark lane. The SMA data
are extended in the vertical direction, coincident with the brighter lobe of the scattered light, 
possibly indicative of additional heating. For the SMA data, the contours start at $\pm$2$\sigma$, 
3$\sigma$ and increase in units of 3$\sigma$. For the CARMA data, the contours start at $\pm$3$\sigma$
 and increase by 3$\sigma$ until 15$\sigma$ where they increase by 5$\sigma$. The RMS noise level
 ($\sigma$) in the 870 \micron\ data is 11 mJy beam$^{-1}$ and 
0.29 mJy beam$^{-1}$ in the 3.4 mm data.}
\label{multiarr}
\end{figure}

The observations of L1527 in multiple configurations enable the physical structure of the envelope
and disk to be examined at multiple spatial scales.
In addition to the most extended configurations, data have also been taken for L1527 in more compact configurations
at both the SMA and CARMA. The resultant images are shown in Figure \ref{multiarr}, overlaid on the L\arcmin\ image of
the edge-on scattered light structure from \citet{tobin2008}. The combined datasets
 trace larger-scale structures, but emphasize different features. The CARMA 3.4 mm data have extended
emission on $\sim$4\arcsec\ scales that could be tracing the extended disk structure or a rotationally-flattened inner envelope 
meeting the disk. The SMA image remains more compact, but trace smaller-scale extended emission 
in the same direction as the CARMA large-scale emission; a difference is that the SMA data appear
more extended along the vertical direction of the disk, coincident with the brighter scattered light
feature. The SMA data does not recover larger-scale emission due to a lack of shorter uv-spacings 
and a more sparsely sampled uv-plane. The measured flux densities from these multi-configuration
data are given in Table 3.

\subsection{High-resolution Images and Disk Size}

\begin{figure}[!ht]
\begin{center}
\includegraphics[angle=-90, scale=0.7]{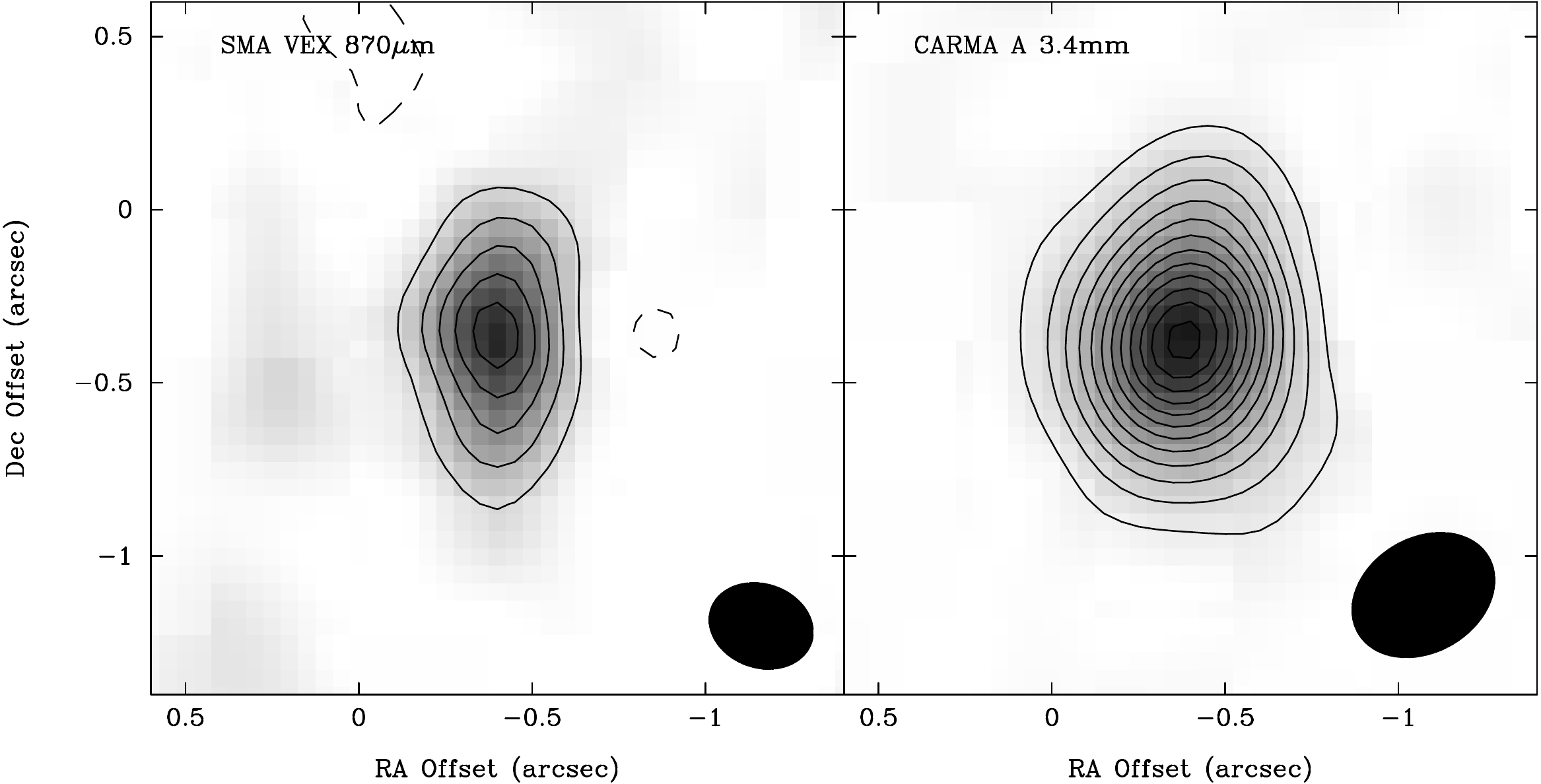}
\end{center}
\caption{High-resolution continuum images for L1527 from the SMA at 870 \micron\ 
(left) and CARMA at 3.4 mm (right), both use natural weighting. 
The contours in all images start at 3$\sigma$ and increase in 3$\sigma$ intervals; $\sigma$ = 5.0 mJy beam$^{-1}$ (SMA),
 and 0.24 mJy beam$^{-1}$ (CARMA 3.4 mm). Both the 870 \micron\ and 3.4 mm data are extended normal to the outflow direction and
in the same direction as the dark lane imaged with Gemini at L\arcmin-band. These data are consistent with L1527 harboring an edge-on disk.
}
\label{carma_sma_avex}
\end{figure}

The naturally-weighted image from the SMA very extended data is shown in the left panel of Figure
\ref{carma_sma_avex}. The image shows dust emission extended
in the direction of the dark lane imaged by Gemini, subtending $\sim$1\farcs0
in diameter. The CARMA A-array 3.4 mm image, also generated using natural-weighting,
is shown in the right panel. In contrast to 
the SMA image, the central point source is more prominent, but there is extension in the 
same direction and with similar angular extent as the SMA data. These
datasets are both tracing the embedded protostellar disk around L1527 and are qualitatively similar
to other observations of edge-on disks surrounding more evolved sources \citep[e.g.][]{wolf2008}.

Since these observations resolve the protostellar disk, one of the key
parameters to measure is its size. From the very extended and A-array data alone, the disk size 
is measured to be $\sim$0\farcs95$\times$0\farcs55 
($\sim$133$\times$77 AU $\pm$ 19 AU) and $\sim$1\farcs15$\times$0\farcs85 ($\sim$160$\times$120 AU $\pm$ 12 AU) in the SMA
and CARMA images respectively. The size was determined by measuring 
the maximum dimension enclosed by the 3$\sigma$ contour along the long 
and short axis of emission in the convolved image. The size of the short axis of the disk will 
be over estimated due to convolution with the beam; however, the major axis length is 
resolved in both images. The full width at half maxima (FWHM) from fitting the images with a two dimensional
Gaussian are approximately half the length measured inside the 3$\sigma$ contour. The deconvolved FWHM are 
comparable to the FWHM of the Gaussian along the major axis, but the deconvolved sizes of the minor
axes are about a factor of two smaller than the fit. Values for the Gaussian fits are given in Table 3.

\begin{figure}
\begin{center}
\includegraphics[angle=-90, scale=0.7]{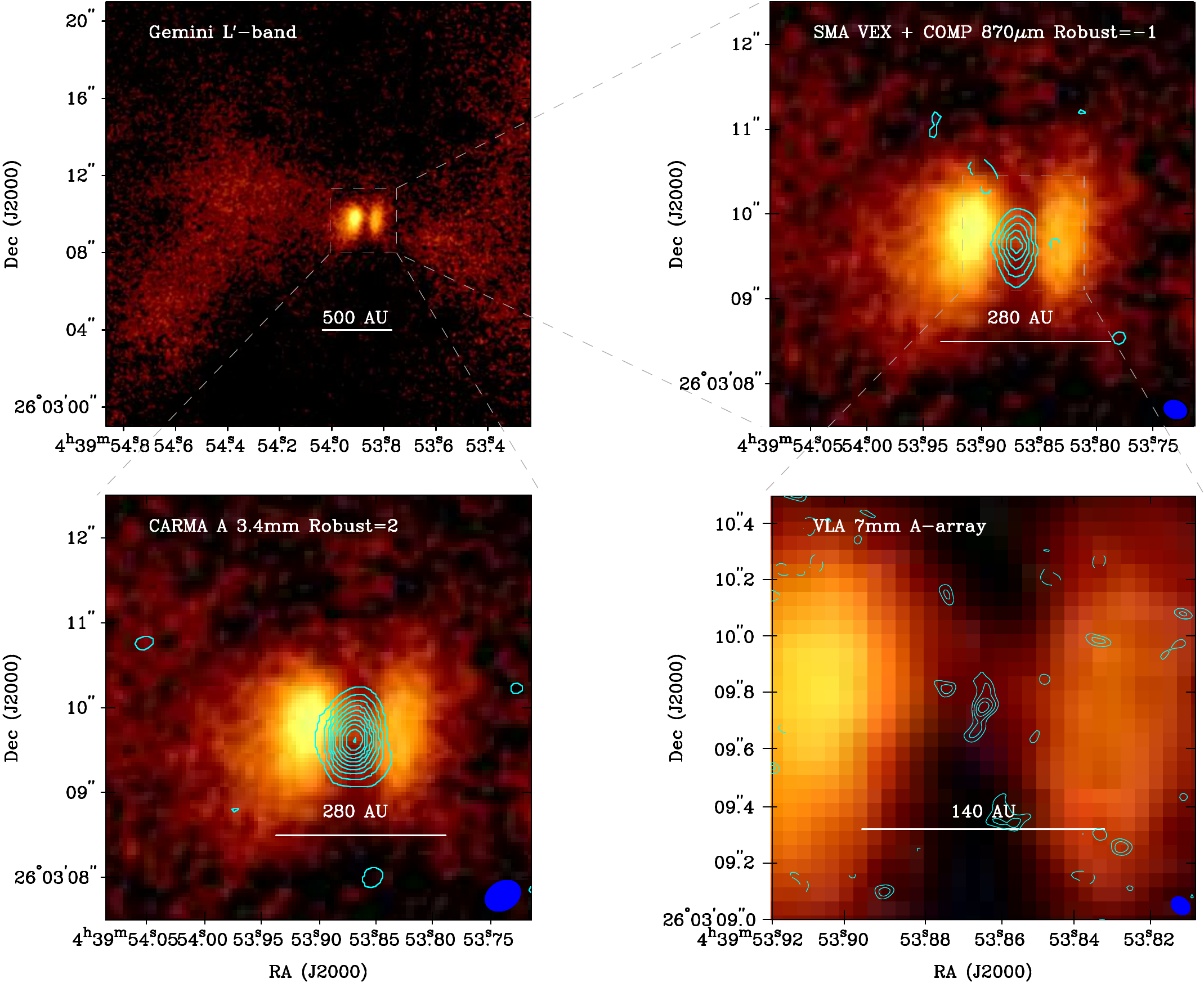}
\end{center}
\caption{L$^{\prime}$ image of L1527 with millimeter continuum data overlaid on the sub-panels.
Top Left: L$^{\prime}$ image of L1527 showing the full range of the high-resolution Gemini data.
Top Right: Zoom-in on the L$^{\prime}$ image with SMA 870 \micron\ data overlaid on the L$^{\prime}$ image.
Bottom Left: CARMA 3.4 mm data overlaid on the zoomed image. Bottom Right: Further zoom-in on the L$^{\prime}$
data with VLA 7mm data from \citet{loinard2002} overlaid. Note that the millimeter data were manually centered
on the dark lane as the registration of the L$^{\prime}$ image is approximate. 
The contours in all images start at 3$\sigma$ and increase in 3$\sigma$ intervals; $\sigma$ = 5.0 mJy beam$^{-1}$ (SMA),
0.24 mJy beam$^{-1}$ (CARMA 3.4 mm), 0.3 mJy beam$^{-1}$ VLA.}
\label{mm-mir}
\end{figure}

The millimeter images are overlaid on the L\arcmin\
image from Paper I in Figure \ref{mm-mir}. The radius of disk inferred from the millimeter emission
is $\sim$ 2x smaller than what was determined from modeling in
Paper I. However, the smaller region of resolved millimeter emission
is not inconsistent with a larger disk being present as the
 emission may simply be too faint to detect and/or be resolved-out
at these large radii.  The 7 mm VLA A-array image
from \citet{loinard2002} is also included in this plot, showing the small-scale
emission from the inner disk; these data are clearly missing substantial
larger-scale flux from the disk due to lack of shorter uv-spacings and/or sensitivity.

\subsection{Visibility Data}
\begin{figure}
\begin{center}
\includegraphics[scale=0.5]{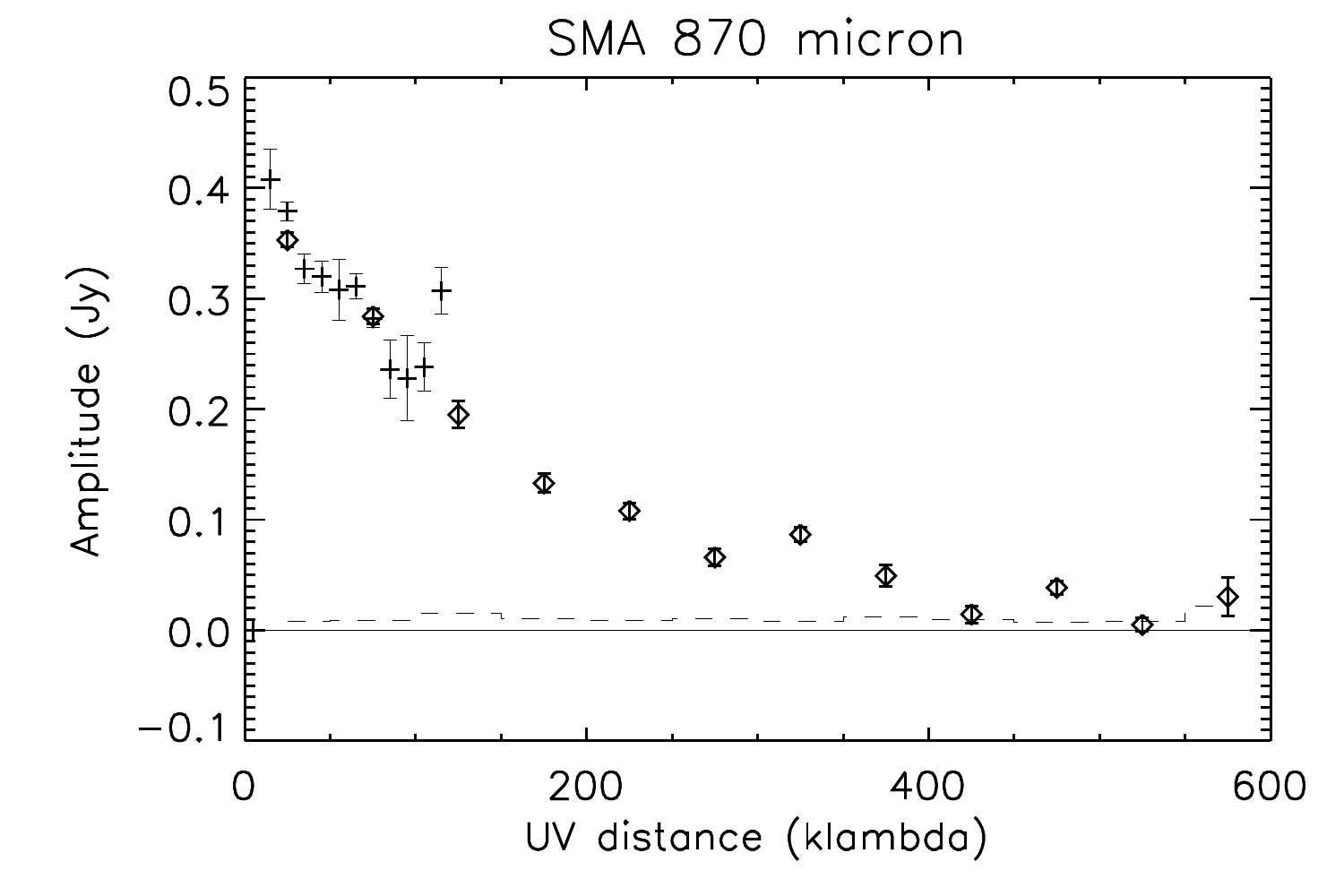}
\includegraphics[scale=0.5]{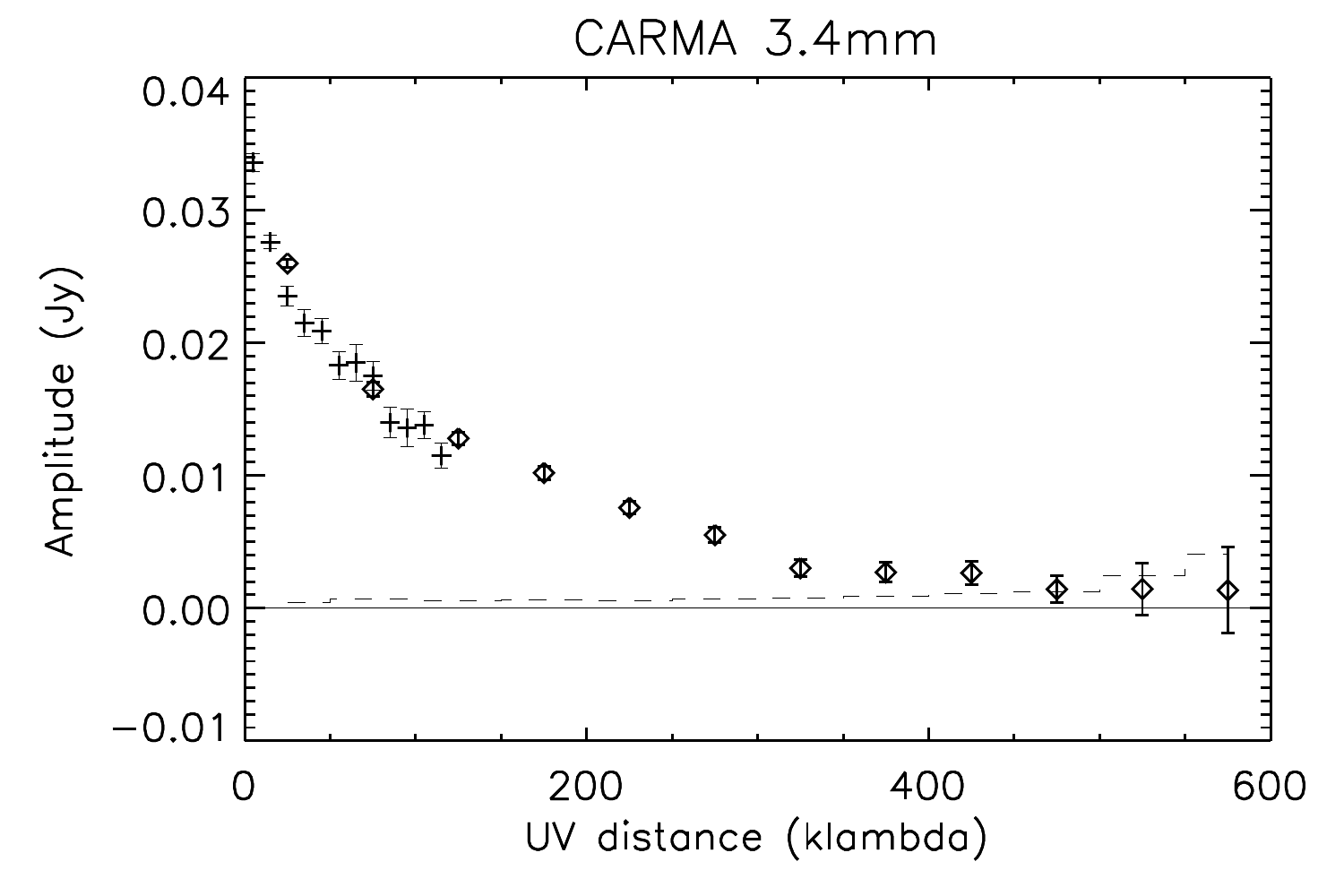}
\includegraphics[scale=0.5]{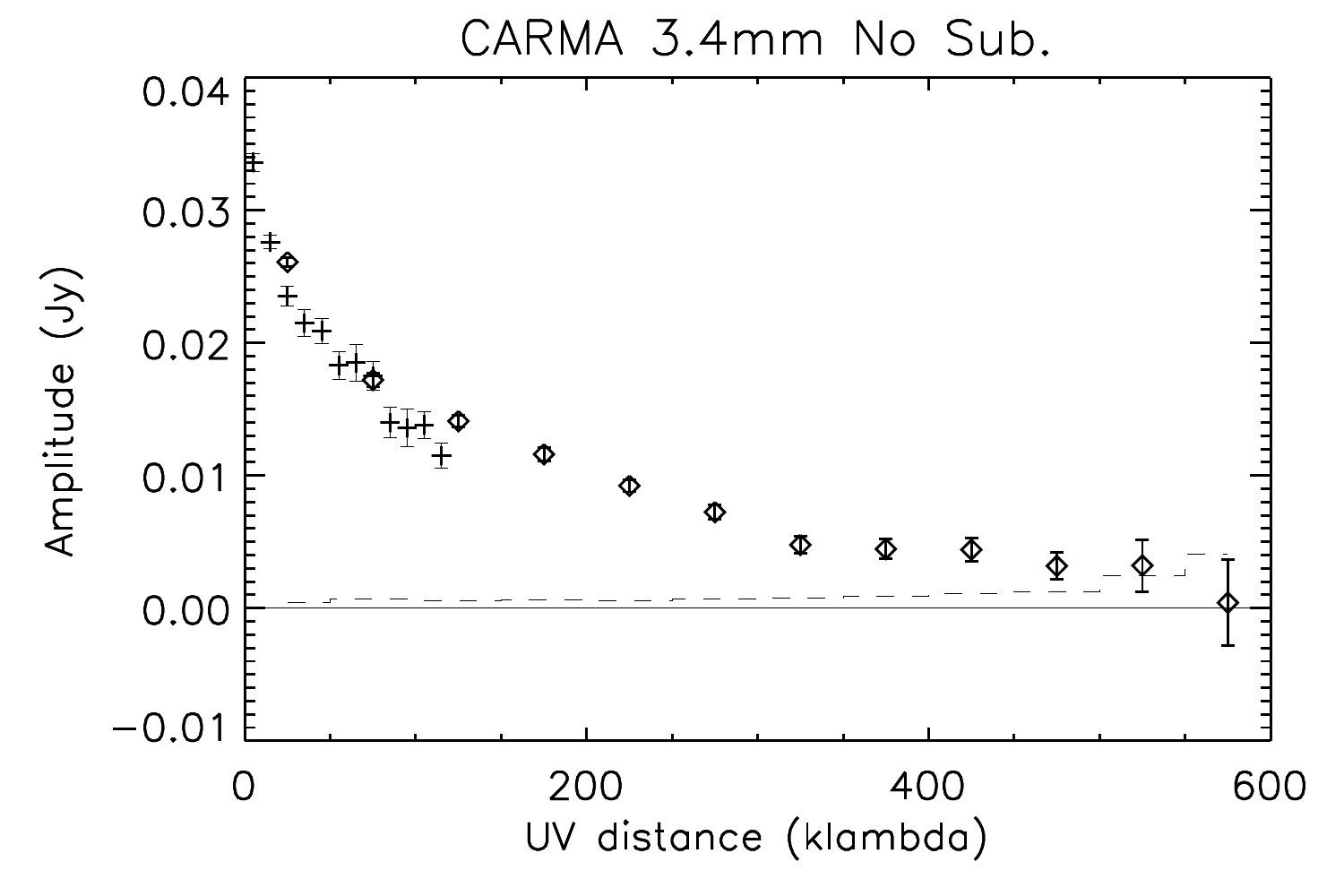}
\end{center}
\caption{Plots of visibility amplitude versus projected baseline for the 
SMA 870 \micron, CARMA 3.4 mm, CARMA 3.4 mm point-source subtracted. 
The different symbols represent different binning of the visibility data in the uv-plane, in 10 k$\lambda$ bins (plus-signs)
and 50 k$\lambda$ bins (diamonds). The relative shapes of the visibility curves in the CARMA and SMA data are similar;
the CARMA 3.4 mm visibilities without the point-source subtracted flatten-out longward of 300k$\lambda$ at $\sim$1.8 mJy.
The fact that the SMA data show no similar flattening further motivates the subtraction of the point-source as contaminating free-free
emission. The uncertainties are only statistical, not including the absolute calibration uncertainty. The dashed line which is near zero
at short uv-distances represents the expected visibility amplitude from a blank field.
}
\label{visibilities}
\end{figure}

We examine the flux density versus uv-distance in
Figure \ref{visibilities}. The visibilities amplitudes from all observed array configurations in the two
wavelengths bands are shown in the individual subpanels. Between uv-distances of 0 to $\sim$200 k$\lambda$
the amplitude vs. uv-distance trend is similar at all wavelengths; however, at uv-distances longward of 300 k$\lambda$
the CARMA 3.4 mm visibilities flatten out. At these uv-distances the SMA data are continuing a downward
trend, but are still above the noise floor. The differences between the CARMA and SMA 
visibilities indicate that the 3.4 mm data have an additional 
unresolved component to the flux that is not present or substantially weaker at 870 \micron.
Flat visibility curves at long baselines in protostellar objects have previously been used to infer
the presence of an unresolved disk \citep[e.g.][]{harvey2003}. However, if we are examining dust emission from
the same structure at both 870 \micron\ and 3.4 mm, then the visibility curves should be nearly identical.

The most likely culprit for the unresolved emission is the thermal jet, radiating free-free emission.
\citet{reipurth2002} observed L1527 with the VLA in A-configuration at 3.6 cm (0\farcs3 resolution),
detecting a point source and \citet{melis2011} conducted new observations with the EVLA between 7 mm and 6 cm 
finding that S$_{\lambda}$(ff) $\propto$ $\lambda^{-0.33}$. Extrapolating this slope to 3.4 mm,
the free-free contribution to the continuum flux is expected to be 1.76 mJy, consistent
with the visibility amplitudes measured at baselines
$>$300 k$\lambda$. Thus, the free-free jet emission is contaminating the dust continuum emission at
3.4 mm. Its removal from the data is trivial since the emission is point-like. We subtracted
the 1.76 mJy point source from all 3.4 mm datasets using the MIRIAD task \textit{uvmodel}
 centered on the continuum peak position at 3.4 mm.

The visibility plot from the subtracted dataset
is shown in Figure \ref{visibilities}. The flux densities measured from the CARMA (free-free corrected) 
3.4 mm A-array and SMA 870 \micron\ very extended images are given in Table 3. With the corrected data,
we can clearly show that the observations toward L1527 do not show the flat visibilities at
 long baselines, due to the extended nature of the disk and emission being resolved across all observed
uv-distances.

\subsection{Millimeter SED}

\begin{figure}
\begin{center}
\includegraphics[scale=0.5]{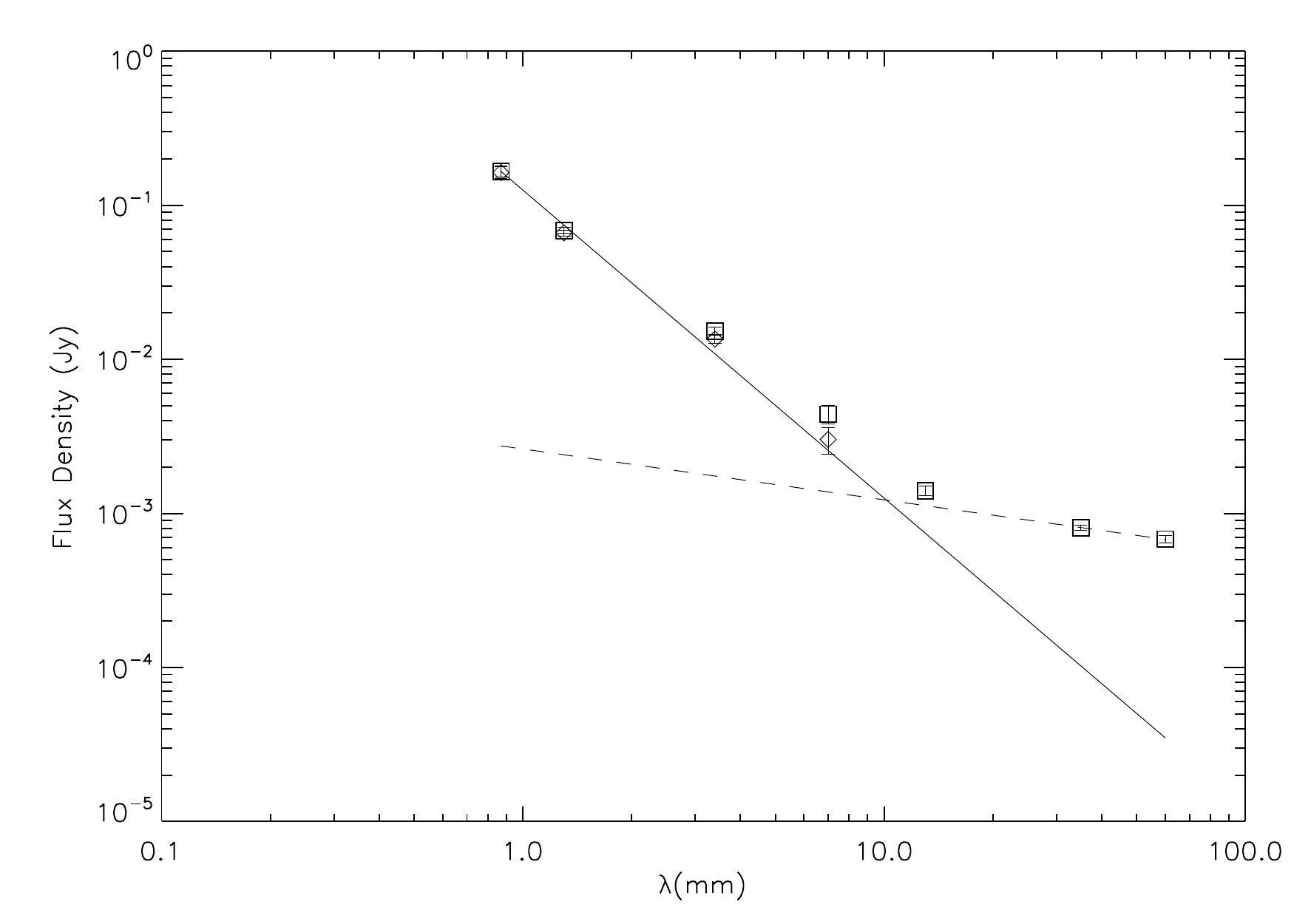}
\end{center}
\caption{SED of L1527 from 870 \micron\ to 6.5 cm; points longward of 3.4 mm are from \citet{melis2011}. 
The 870 \micron\ and 3.4 mm points are measured
from the naturally weighted images in the very extended and CARMA A configurations respectively. The dashed line
is the fit to the free-free jet emission in the centimeter \citep{melis2011} with a spectral index 
of 0.33; the solid line is a power law
with a spectral index of 2 (Rayleigh-Jeans). The square points are the measured photometry
 without correction for free-free emission and
the diamonds have the free-free emission subtracted. The error bars are only statistical uncertainties, not including the $\sim$10\% error
in flux calibration.}
\label{mm_sed}
\end{figure}

In the millimeter regime, if the dust temperature is isothermal and optically thin, the spectral index of
thermal dust emission should follow the relation
\begin{equation}
F_{\lambda} \propto F_{\lambda,0}\left(\frac{\lambda_0}{\lambda}\right)^{\alpha},
\end{equation}
with $\alpha$ = 2 + $\beta$, assuming the Rayleigh-Jeans limit, and $\beta$ is 
the dust opacity spectral index. Therefore, if the emission is 
optically thin and isothermal we can calculate $\beta$ directly from the sub/millimeter flux densities.
The flux densities measured at CARMA A-array and SMA very extended array yield $\alpha$ $\sim$ 1.9, implying
$\beta$ $\sim$ 0. We plot our observed fluxes at 870~\micron\ and 3.4~mm along with 1.3~mm fluxes from \citet{maury2010} and 
the 7~mm to 6~cm fluxes observed by \citet{melis2011} in Figure \ref{mm_sed}. All these data are consistent 
with $\alpha$ = 2 and $\beta$ = 0, given the assumptions of isothermal and optically thin emission.
In the absence of large flux calibration offsets, this result could indicate the following: 
the emission is becoming optically thick, the dust is not isothermal, and/or a gray opacity law in the millimeter. 
A gray opacity would result if the dust grains probed by these observations were larger than the
observed wavelength. However, \citet{melis2011} suggested that their 1.3~cm data indicated a lack of cm-sized
grains, but \citet{scaife2012} used a different fitting method and had more cm-wave data that
indicated a presence of cm-sized grains, see section 5 for further discussion.

\begin{figure}
\begin{center}
\includegraphics[scale=0.6]{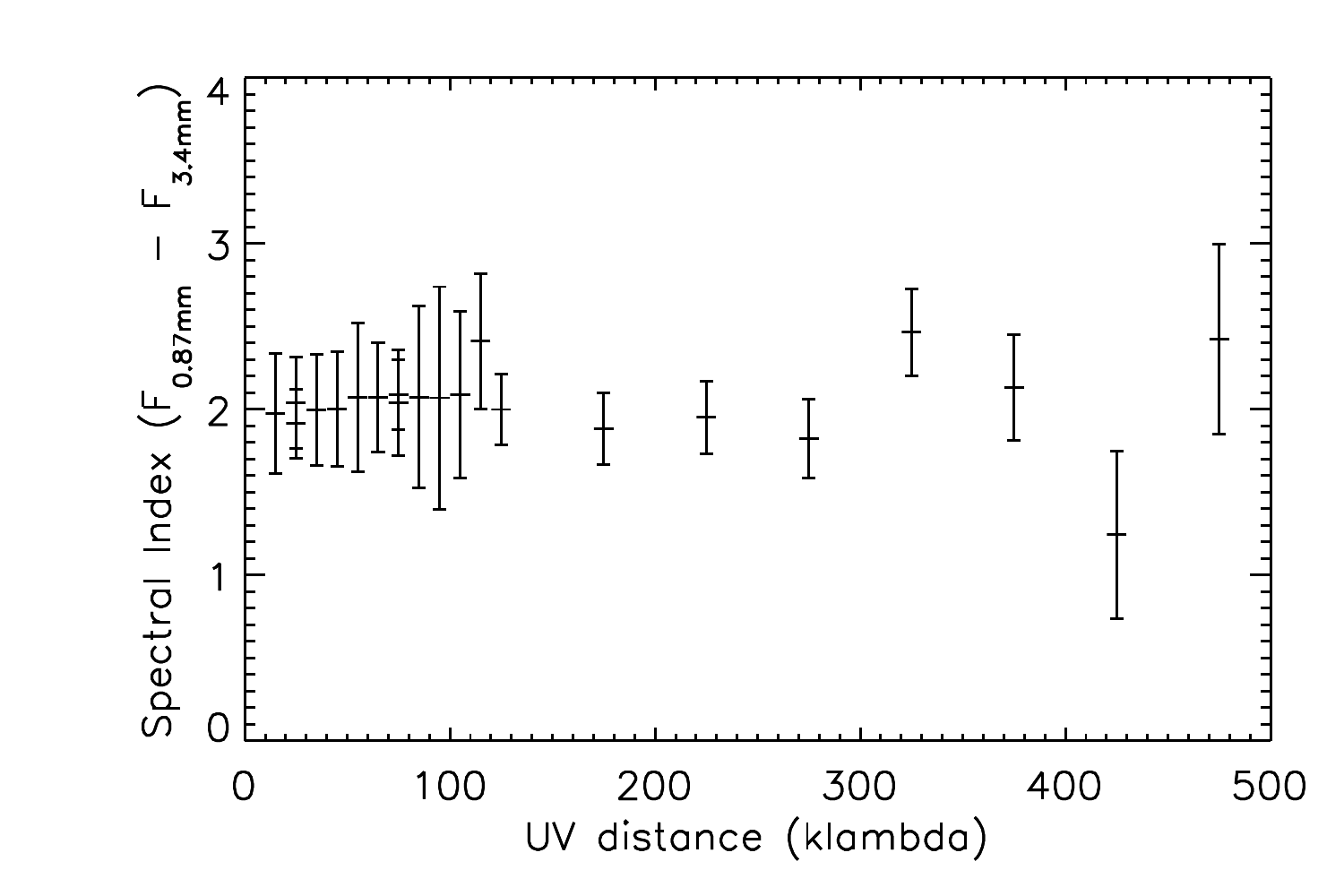}
\end{center}
\caption{Spectral index of emission between 870 \micron\ and 3.4 mm emission. Assuming optically thin emission, the
spectral index is equal to 2 + $\beta$ in the Rayleigh-Jeans limit. The L1527 data are consistent with $\beta$ $\sim$ 0
across all uv distances. The errorbars include both statistical uncertainty and 20\% flux calibration uncertainty at each wavelength.}
\label{alpha_uvd}
\end{figure}

We have examined the spectral index of the visibility amplitudes versus
uv-distance in Figure \ref{alpha_uvd} (using the free-free corrected data at 3.4 mm). 
Across the overlapping uv-range the emission is consistent with
$\alpha$ = 2 at both large (envelope; 1000s of AU) and small (disk; 100s of AU) spatial scales.\
A caveat here could be that the flux densities are only accurate to $\sim$10\% at best, 
which yields an uncertainty in $\beta$ of $\sim$0.1 \citep{chiang2012}; for 20\% errors in flux calibration
the uncertainty increases to 0.21. Therefore, even with large amplitude calibration errors the shallow spectral index
of the disk appears to be real. A shallow opacity spectral index is not necessarily unexpected given that
Taurus Class II sources also often consistent with $\beta$ $\sim$ 0 using optically thin assumptions \citep{andrews2005,ricci2010}.
In addition, \citet{kwon2009} found shallow $\beta$ values (0.5 - 1.0) in protostellar envelopes on 1000s of AU scales.

\subsection{Disk Mass}
The disk mass can be simplistically calculated using the same assumptions of isothermal and optically
thin dust emission as we used to determine the spectral index.
We assume a dust opacity law normalized to $\kappa_0$=3.5 cm$^{2}$ g$^{-1}$ (dust only) at 850\micron \citep{andrews2005}, a 
dust-to-gas mass ratio of 1:100, and the observed dust opacity spectral index ($\beta$) = 0.
 We then calculate the mass assuming optically thin emission and constant dust temperature
with the following equation
\begin{equation}
M_{dust} = \frac{D^2 F_{\lambda} }{ \kappa_0\left(\frac{ \lambda }{ 850\mu m }\right)^{\beta}B_{\lambda}(T_{dust}) },
\end{equation}
where $D$ = 140 pc and $T_{dust}$ is estimated to be 30K. 
The flux densities at 870 \mum\ and 3.4 mm
were taken from Table 3, using the SMA very extended and CARMA A-array measurements.
This yields mass estimates of 0.007 and 0.007 $M_{\sun}$ at 3.4 mm and 870 \micron\ respectively, with a statistical uncertainty of
$\pm$ 0.0007 $M_{\sun}$. If we had not assumed $\beta$ = 0, and chosen $\beta$ = 1 as is typically done \citep{andrews2005,andrews2009},
then there would be a factor of $\sim$3 discrepancy in the mass calculated at different wavelengths.
The possible reasons for such a shallow $\beta$ in this source are further discussed in Section 5.

\section{Modeling}

To better understand the physical properties of the disk around L1527, we model the thermal dust emission
in the sub/millimeter, L\arcmin\ image, and multi-wavelength SED.
We use the Monte Carlo radiative
transfer code of \citet{whitney2003} to calculate the propagation of radiation through the disk and envelope,
determining the spectral energy distribution (SED) from the near-infrared to the millimeter, producing
scattered light and thermal images in the near to far-infrared; external heating is also considered by
the standard interstellar radiation field extincted by A$_V$ = 3. We use the envelope density structure
from the rotating collapse model \citep[][hereafter CMU model]{ulrich1976, cassen1981, tsc1984}.
This density structure is spherical at large radii, but near the centrifugal radius ($R_C$) 
the envelope becomes flattened due to rotation; the envelope density scaling is linked to
the mass infall rate. Outflow cavities are also included in this modeling with
the same parameters as in Paper I. The density structure of the disk is defined by a radial density profile, 
flaring as a function of disk radius, Gaussian vertical density profile, an initial scale height, and total mass.

We have run a grid of
models, keeping the envelope properties constant, while varying the properties of the 
circumstellar disk (radius, flaring, radial density profile, scale height, and mass). The central
protostellar mass is assumed to be 0.5 $M_{\sun}$\footnotemark\footnotetext{Models were 
run before obtaining the \citet{tobin2012} protostellar mass measurement.}; however, no parameters directly depend 
on this value. The protostar mass is only used to calculate the disk-protostar accretion 
luminosity and convert the envelope mass infall rate into 
the envelope density of the CMU model. The mass infall rate is taken to be 1.0 $\times$ 10$^{-5}$ $M_{\sun}$ yr$^{-1}$,
identical to \citet{tobin2008}, but 25\% larger than that used in Paper I. We also assume the same disk-protostar accretion
rate, yielding L$_{acc}$=1.75 $L_{\sun}$ and a protostellar luminosity of 1.0 $L_{\sun}$. Note that 
the measured protostellar mass of 0.19 $M_{\sun}$ suggests that the luminosity of the protostar 
may be lower and the accretion luminosity would be greater;
however the source of emission will not significantly alter the sub/millimeter emission. Table 4 gives a complete
list of model parameters and identifies those varied or fixed. The total number of models run was 3,584, using five free parameters.

The key constraints on the disk properties for L1527 lie in 
the high-resolution sub/millimeter and L\arcmin\ images. 
The \citet{whitney2003} code can produce images at these wavelengths; however,
the images are constructed from the individual photons that are either absorbed/reemitted or scattered
on their way out of the envelope. This imaging method works well for scattered light at L\arcmin-band and
other wavelengths where there is substantial flux, but an impractical number of input photons 
are needed to produce reasonable high-resolution sub/millimeter 
images. Therefore, to generate high-resolution sub/millimeter images with high-signal to noise, we use the temperature
and density structure calculated by the \citet{whitney2003} code as input to 
the LIME (LIne Modeling Engine) radiative transfer code \citep{brinch2010},
using the ray-tracing function to generate dust continuum images.
The MIRIAD task \textit{uvmodel} is used to calculate the Fourier transform of the model image and sample it with the
same uv-coverage as the CARMA and SMA observations. We then use the \textit{uvamp} task to
calculate the flux versus projected baseline for a given model in circularly averaged bins of uv-distance
identical, to those in Figure \ref{visibilities}. Note that while the uv-coverage is identical,
the effects of atmospheric decorrelation and thermal noise are not taken into account for the model visibilities.

\subsection{Dust Opacities}

The dust model used for the envelope and disk with the \citet{whitney2003} code 
is the same as presented in \citet{tobin2008}, with grains up to $\sim$1~\micron\ in radius
and an assumed mixture of discrete spherical grains composed of graphite, astronomical silicates, and 
water ice. The dust opacity spectral index ($\beta$) of this
model at sub/millimeter wavelengths is 1.93; ISM dust grains have $\beta$ $\sim$ 2 for comparison.
However, we found that these dust opacities were found to be too low 
(0.6 cm$^{2}$ g$^{-1}$ at 870 \micron; opacity of dust only) at sub/millimeter wavelengths. Moreover, the spectral index of these
opacities is much steeper than found in the observations, 1.93 versus $\sim$0. Therefore, models of dust emission
using the same opacities from the radiative equilibrium code cannot simultaneously reproduce the emission
at both 870 \micron\ and 3.4 mm. It is possible that the disk is simply very massive and optical depth
effects are causing the the shallow observed spectral index; however, then the disk would be too opaque to 
reproduce the L\arcmin\ scattered light structure. We also cannot simply use larger-grain models in the
radiative equilibrium calculation because that makes the SED fit bad from $\sim$3 \micron\ to 70 \micron,
due to the sensitivity of the mid-infrared SED shape to the maximum grain size.

A dust model with maximum grain sizes between $\sim$1 to 10 cm and a shallow 
power-law size distribution $N(a)$ $\propto$ $a^{-2.5}$ - $a^{-3.0}$
could reproduce the observed spectral index from 870 \micron\ to 3.4 mm 
\citep{dalessio2001,hartmann2008,ricci2010} and \citet{scaife2012} indicated that there
was evidence for cm-sized grains in L1527.
Given these uncertainties, we chose to adopt a parametrized sub/millimeter 
dust opacity in the LIME raytracing calculation to simplify modeling because a detailed
exploration of dust opacities is beyond the scope of this study. 
Other published studies \citep[e.g.][]{looney2003,chiang2012} have adopted similarly parametrized
sub/millimeter dust properties, deviating from the dust opacity model used at shorter wavelengths in the radiative
transfer calculation. The need to adopt a parametrized dust opacity at sub/millimeter wavelengths reflects the
inability of dust models to fully describe the opacities across all wavelengths with a single self-consistently generated
model or spatial variation in the dust size distribution.

We normalized the sub/millimeter dust opacity at 850 \micron\ to be
of 3.5 cm$^{2}$ g$^{-1}$ \citep{andrews2005} and varied 
the spectral index between 0.0 and 0.75 in steps of 0.25.
The use of a different opacity in the millimeter should not impact the calculated
dust temperature structure, because most energy is absorbed by the
dust grains at wavelengths shortward of 100 \micron.
The small values of the dust opacity spectral index 
modeled are motivated by the shallow spectral index from 3.4 mm to 870 \micron\ and that other 
millimeter studies of protostars often find values of $\beta$ $\la$ 1 \citep{kwon2009,chiang2012}.
 Including these dust opacity variations, the total number of models increases to 14,336.

\subsection{Model Fitting}

The large number of models generated with combinations of the 
five parameters explored necessitates using statistical tests to 
weed out the large fraction of models that do not fit the data. To do this, we simultaneously fit the
870 \micron\ and 3.4 mm visibilities, sub/millimeter images, the multiwavelength SED, and IRS spectrum by
calculating individual $\chi^2$ values for each type of data and computing the average.

We calculate the $\chi^2$ of the azimuthally averaged model visibilities 
at 870 \micron, and 3.4 mm using the equation
\begin{equation}
\chi^2 = \sum_{i}^{N}\frac{(F_{\nu,observed,i} - F_{\nu,model,i})^2}{\sigma_i^2}
\end{equation}
for the nine visibility points between 0 and 400 k$\lambda$ taken in 50 k$\lambda$ bins, as shown
in Figure \ref{visibilities}. The uncertainty in the data, $\sigma_i$, includes
the statistical uncertainty in addition to a 5\% absolute flux uncertainty; although the true
absolute flux uncertainty for these data is $\sim$10\%, we could not assign such an uncertainty to each
point without making many models appear overly well-constrained. This is because the absolute
flux uncertainty is an overall offset in the scaling of the flux densities and not random noise. Also note
that once the signal to noise of the visibility data in a given bin is less than $\sim$3, the errors are no longer
Gaussian but follow the Rice distribution. We do not expect this to significantly affect our fitting as the lowest
signal-to-noise is 3.7 for one point in the CARMA 3.4~mm data; outside our chosen range for fitting the
visibilities do have signal-to-noises less than 3. 

We calculate the $\chi^2$ of the spectral energy distribution from the near-infrared to the 
millimeter, in a similar manner to equation 3; this to ensures that the models fitting the millimeter 
visibilities also result in a reasonable SED. The IRS spectrum is also fit from 11 $\micron$
to 20.7 \micron. We select a subset of the full IRS spectral range in order to avoid ice
features present between 6 \micron\ and 11 \micron\ as well as the wings of the 45 \micron\ water ice
feature apparent at the long wavelength end of the IRS spectrum.

Since the azimuthally averaged visibilities do not contain information about the two-dimensional structure of the source,
we also use the sub/millimeter images as a constraint on the resolved disk structure.
We directly compare the models to the data by subtracting the model
visibilities from the data and then reconstructing the images from the residual uv dataset. We only use the highest resolution
dataset (A-array for CARMA and very extended for the SMA) for the image comparison.
We then determine the maximum residual (positive or negative) in the sub/millimeter images, 
and calculate 
\begin{equation}
 \chi^2_{image} = \frac{Max(residual)^2}{\sigma_{image}^2}
\end{equation}
 as a measure of how well the dust continuum images are fit by the models. 
Future modeling will incorporate the full set of two-dimensional visibility data to constrain
the structure without an imaging step.

For the L\arcmin\ image, we calculate a $\chi^2$ value for an intensity profile taken in a one dimensional cut 0.\farcs5 wide
along the rotation axis of the disk. Such a cut captures the dark lane and emission from the upper layers of the disk. The
$\chi^2$ value is calculated in a similar manner to equation 3.

\subsection{Distributions of Likely Parameters}

An important caveat to the $\chi^2$ analysis is that our $\chi^2$ values cannot be related back to the
probability distribution function. This is because the $\chi^2$ is only valid (in a statistical sense) when the
model is known to be an accurate representation to the data, in our case we are limited by model mis-specification. 
With so many observational constraints, we are very much in the limit of the analytic prescriptions of the disk and envelope
structure being wrong at some level. Therefore, the overall $\chi^2$ value
is simply telling us the models that are the least wrong, not a ``best fit''. 
While a uniquely constrained model is not currently possible,
the results from modeling are suggestive to the properties of the system.

We have listed the parameter ranges of acceptable models using 
the $\chi^2$ -$\chi^2_{best}$ $<$ 3 criterion used by \citet{rob2007} in Table 6. While this is
not a statistically robust measure, it does show a range of models that fit the data reasonably well. 
We attempt to show how the different
sets of data constrain the model parameters in different ways, showing
how the combined datasets yield a more robust selection
of models. We will do this by weighting the $\chi^2$ value by a factor of 10.0 
for combinations of the visibilities, sub/mm images, SED, IRS spectrum,
and L\arcmin\ images; the data not being weighted by 10.0 are still included
in the fit but with weighting factors of 1.0.
This is similar to the method used by \citet{eisner2012}. For
each set of weighted fits, we give the best fitting parameter and
the minimum and maximum values for the set of models that fulfill the $\chi^2$ -$\chi^2_{best}$ $<$ 3 criterion.

The results of the weighted $\chi^2$ and $\chi^2$ -$\chi^2_{best}$ $<$ 3 model analysis are given in 
Table 6. Again we emphasize that these are not statistical fits, but show the range of parameters
that are capable of fitting the data and which set of data best constrains each parameter of the model 
grid. It is noteworthy that no matter which data are 
being weighted, they all have the same lower-limit
on disk radius of 100 AU (the next lowest radius in the grid was 50 AU).
The sub/millimeter data (visibilities and image) yield constraints on the disk mass, radius, dust opacity 
spectral index ($\beta$), and the radial density profile. The sub/mm images provide the tightest constraint
on the disk radius and the sub/mm images combined with the L\arcmin\ image provide
tight constraints on the disk mass and radial density profile. The L\arcmin\ image alone provides
the only constraint on the flaring of the disk; the full two-dimensional visibility data 
set may also provide some constraints for the vertical density structure, which
are lacking in the one-dimensional visibilities. In summary, the range of model fits
underscore the need for spatially resolved imaging to construct accurate physical models of
disks in protostellar systems. Overall, the tightest constraints
on parameters are given by the L\arcmin\ and sub/mm images and
the visibilities; the SED and IRS spectrum complement the images by
ruling out certain models that could reproduce the images but not the overall emission.

\subsection{Representative Model}

Despite the perils associated with $\chi^2$ fitting, the best fit model and most of the 
`acceptable' parameter range does fit the data well `by eye'.
The best fitting disk model has a radius of 125 AU; this is smaller but not inconsistent with 
our value from Paper I. However, the radius cannot be much smaller without diverging from the L\arcmin\ image.
The disk is found to be highly flared, with 
$H$ $\propto$ $R^{1.3}$ and a scale height of 48 AU at 100 AU (H$_{100}$), 
a Gaussian vertical disk structure is assumed. The radial density profile of the disk is found
to be $\rho$ $\propto$ $R^{-2.5}$, the similar to the model in Paper I.
The disk mass is found to be $M_{d}$ = 0.0075 $M_{\sun}$, somewhat larger than 
the 0.005 $M_{\sun}$ value in Paper I, but consistent with the estimate from the sub/mm flux alone.
As noted earlier, the observations were consistent 
with a dust opacity spectral index $\beta$ $\sim$ 0. Accounting for opacity effects in the models, 
we still find $\beta$ to be low, with $\beta$ = 0.25 as the best fit. Using the best
fitting disk model, we calculate that the optical depth of emission $\tau$ = 0.14 at 870 \micron\ and $\tau$ = 0.10 at 3.4 mm.
Therefore, optical depth does not seem to be the cause of the low values of $\beta$.
We will discuss this further in Section 5.3 and 5.4.

\begin{figure}
\begin{center}
\includegraphics[scale=0.5]{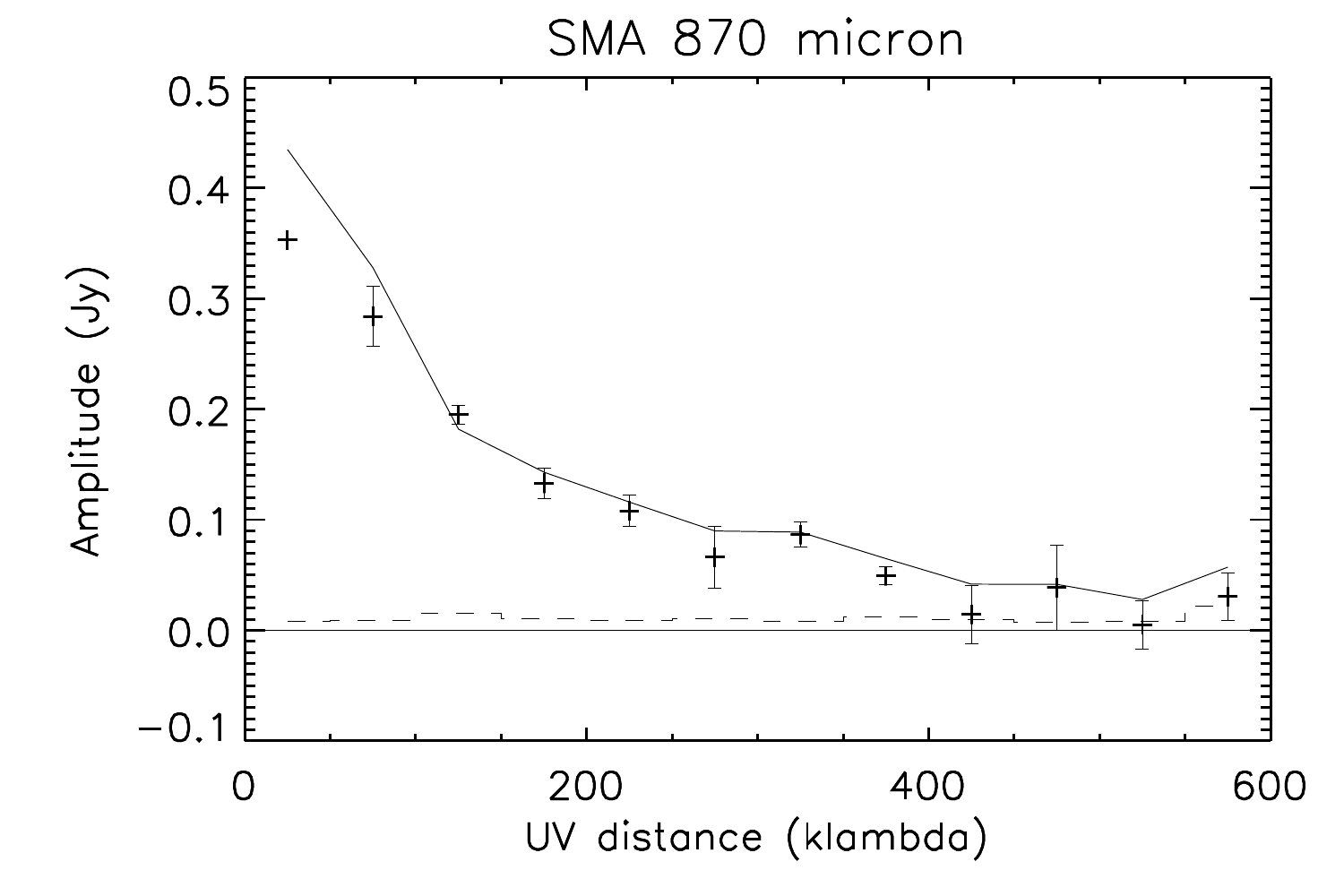}
\includegraphics[scale=0.5]{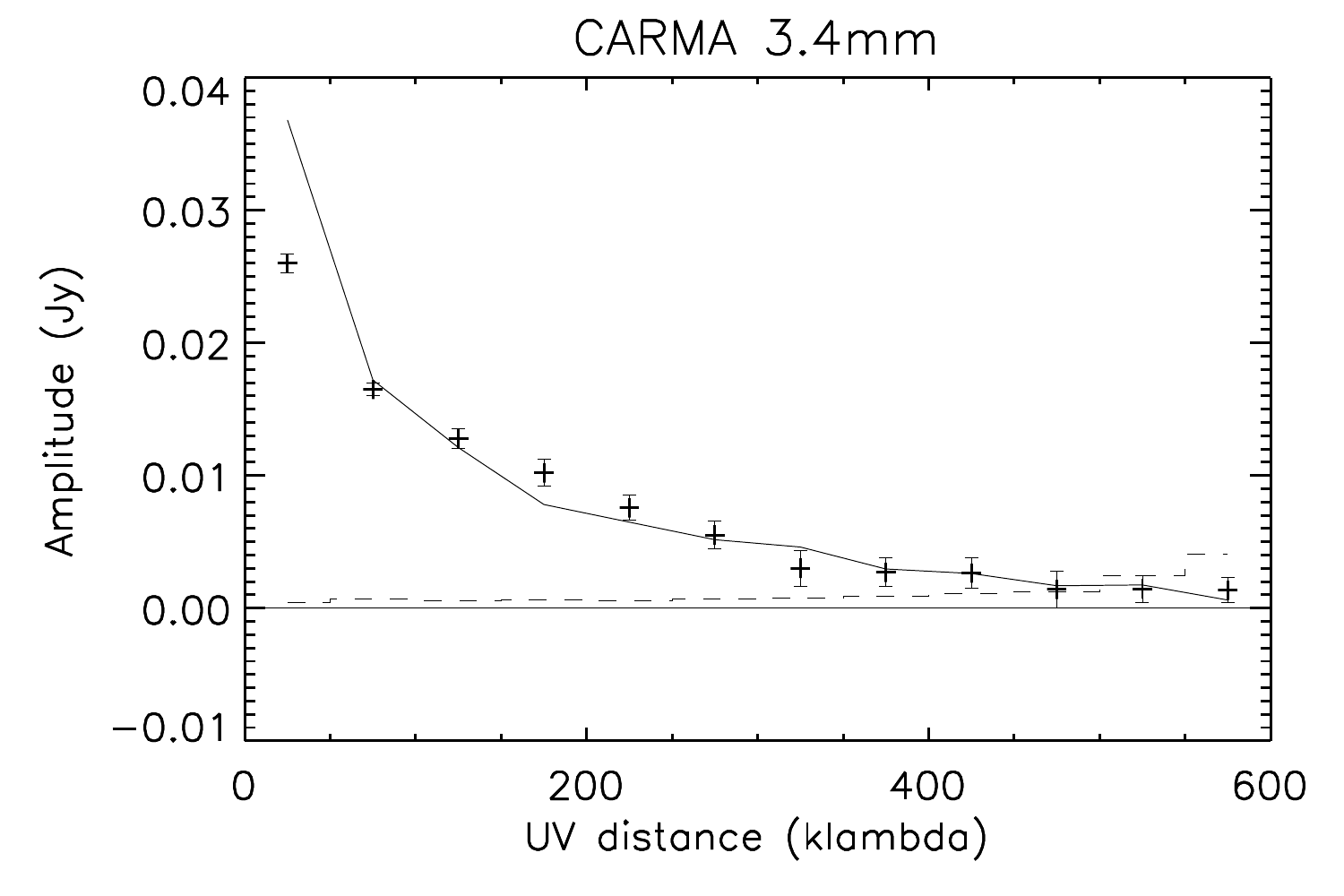}
\end{center}
\caption{
Visibilities of L1527 at 870 \micron\ (left) and 3.4 mm (right) are plotted as the points (same as Figure \ref{visibilities}. The lines
represent the best overall model reproducing the visibility data. 
}
\label{modelvis}
\end{figure}

\begin{figure}
\begin{center}
\includegraphics[angle=-90, scale=0.5]{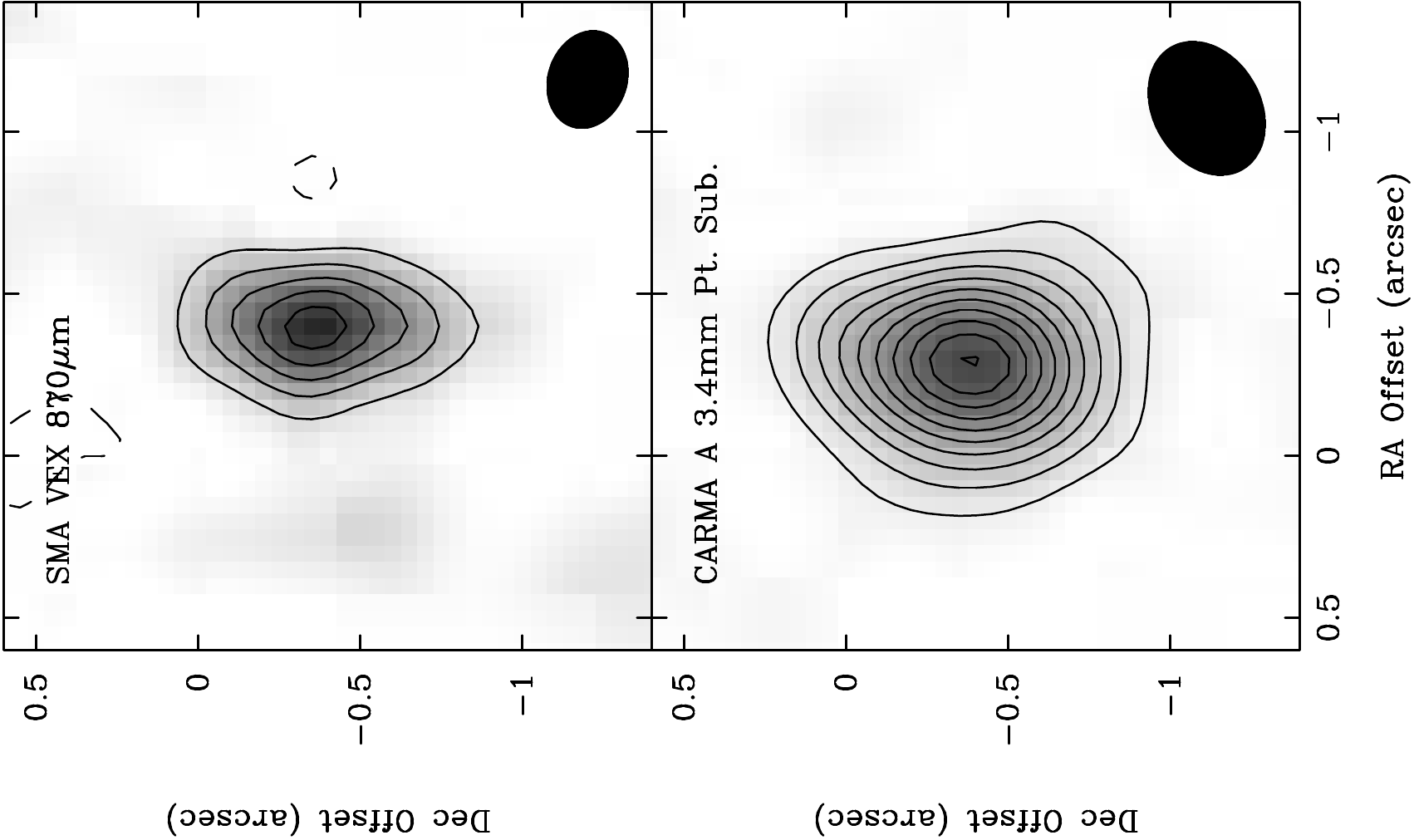}
\includegraphics[angle=-90, scale=0.5]{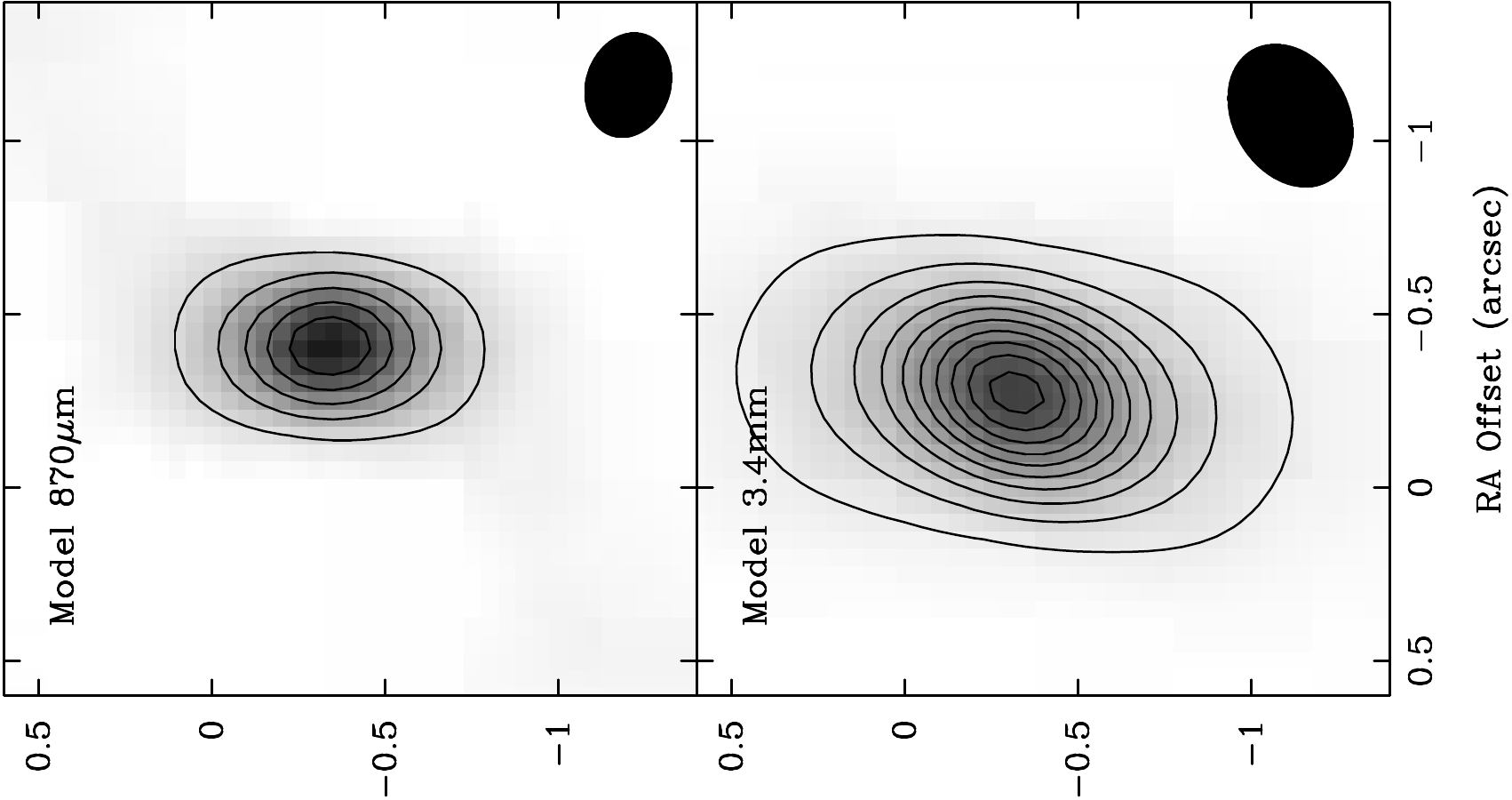}
\includegraphics[angle=-90, scale=0.5]{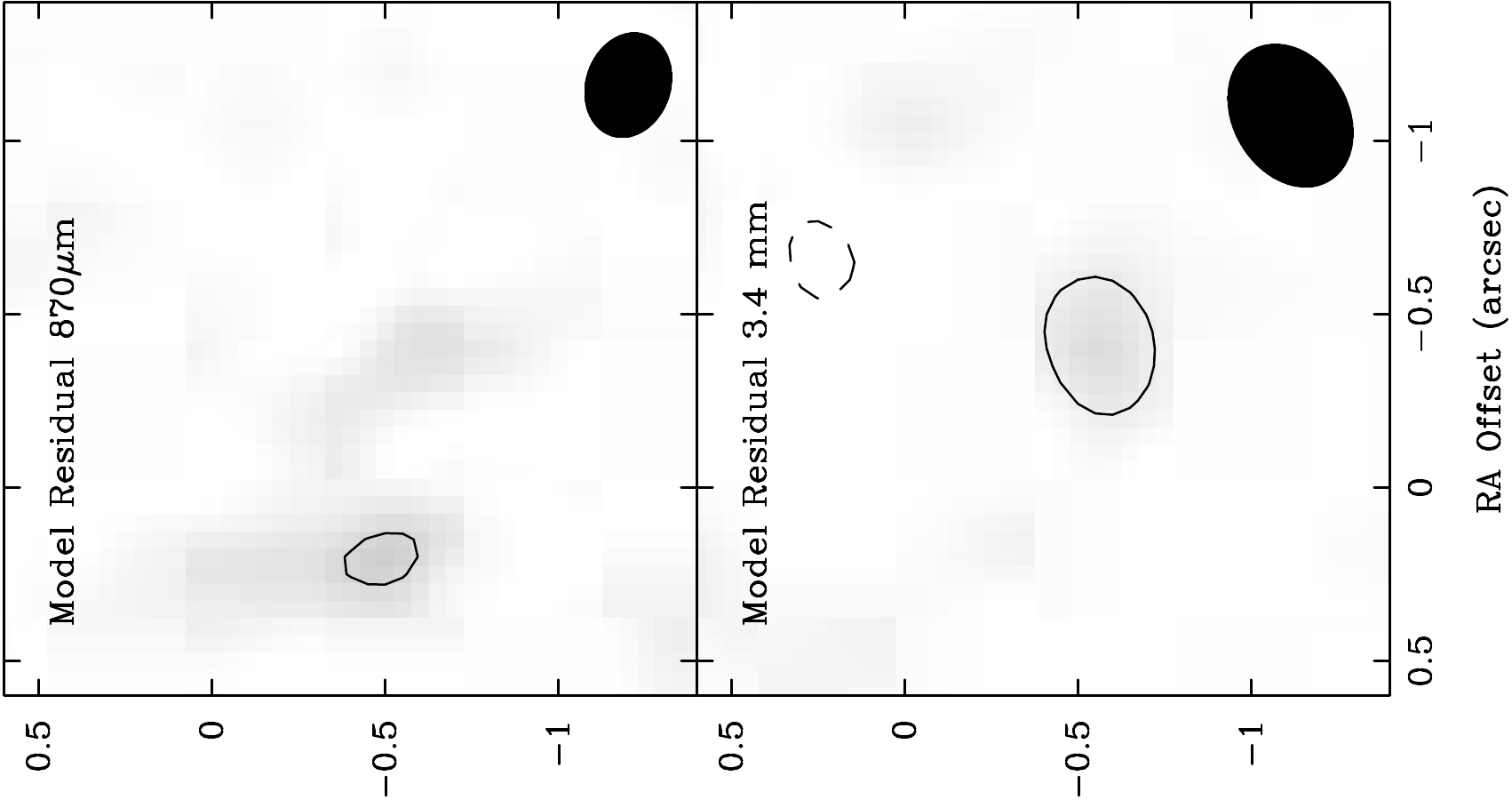}
\end{center}
\caption{Model comparison: Images of L1527 (left) at 870 \micron\ (top) and 3.4 mm (bottom) are shown, 
compared to models at the same wavelength (middle) and the residuals after subtracting
the models from the data in the uv-plane (right). The models capture
the general features seen in the high-resolution data. The most obvious discrepancy
is the 3$\sigma$ residual on the southern side of the disk in the 3.4 mm image. The residual is also 
present in the SMA data, but at a lower level; this appears
to be a disk asymmetry not captured by the model. The contours in all images start at 3$\sigma$ and 
increase in 3$\sigma$ intervals; $\sigma$ = 5.0 mJy beam$^{-1}$ (SMA), 0.24 mJy beam$^{-1}$ (CARMA 3.4 mm).}
\label{modelimage}
\end{figure}

\begin{figure}
\begin{center}
\includegraphics[scale=0.5,angle=-90]{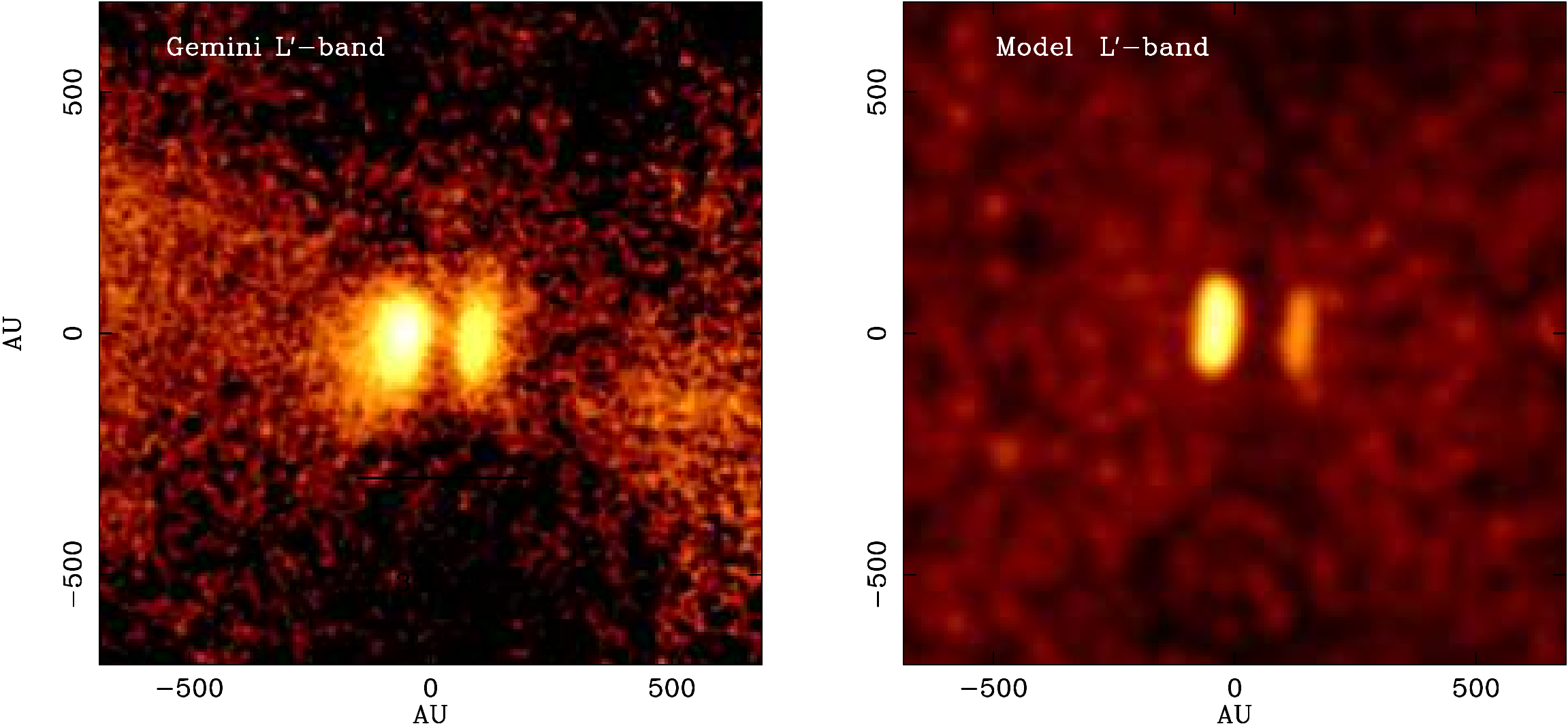}
\end{center}
\caption{Observed L\arcmin\ images of L1527 (left) compared to 
the best fitting model (right); the model
is convolved with a Gaussian with FWHM of 0\farcs35 to match the
seeing of the L\arcmin\ observations. The scattered light
from the model captures the brightest scattered light features of 
the disk well and reproduces the dark lane thickness.
}
\label{modelLimages}
\end{figure}

\begin{figure}
\begin{center}
\includegraphics[scale=1.0]{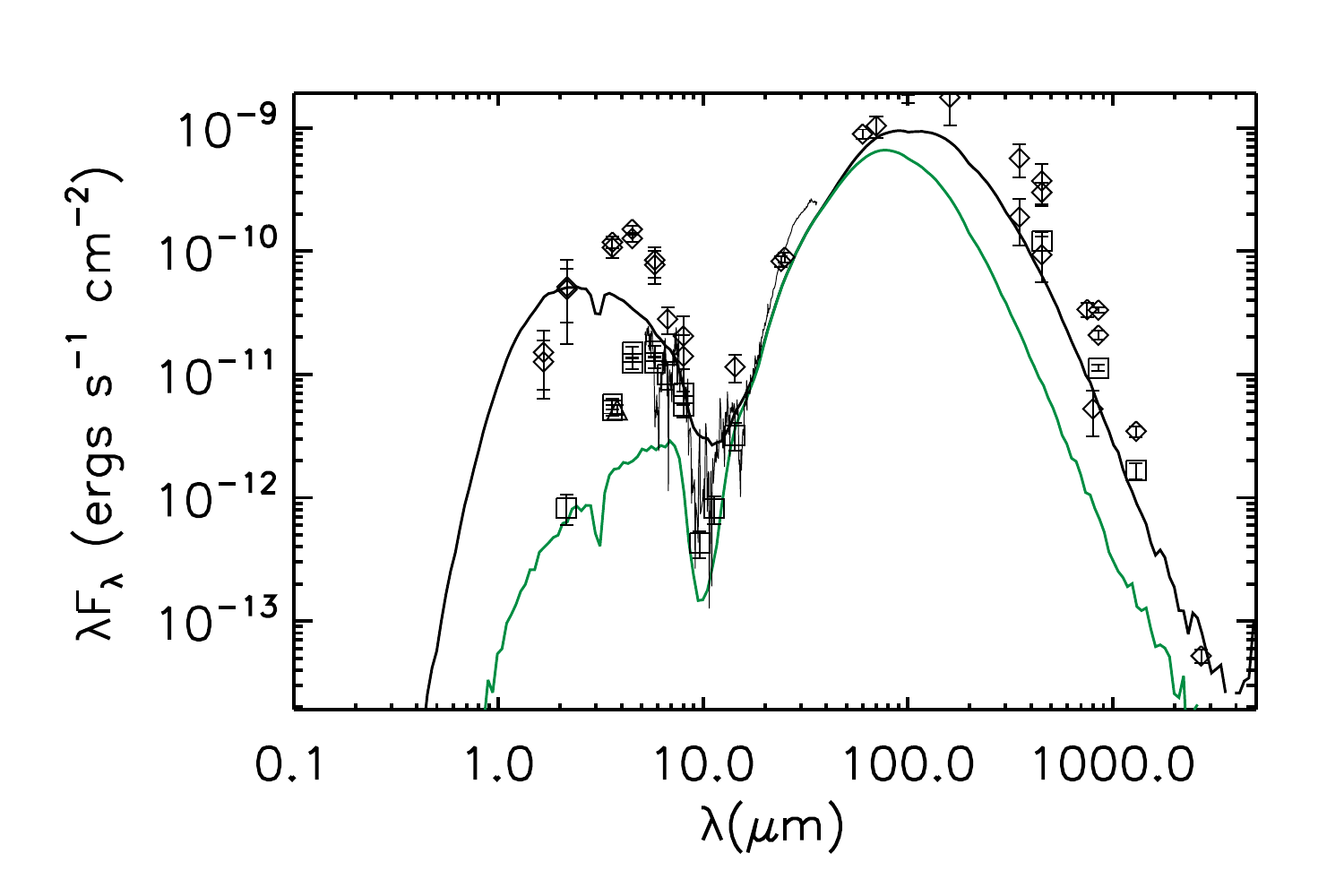}
\end{center}
\caption{Model SED generated with the \citet{whitney2003} radiative transfer code overlaid
on the photometry data.
The photometry and IRS spectrum are taken from \citet[][and references therein.]{tobin2008}.
Photometry taken with apertures of 71\farcs4 (diamonds) and 7\farcs14
(boxes) (10000 AU and 1000 AU) are plotted. The triangle
at 3.8 \micron\ is the Gemini L\arcmin\ flux within 1000 AU.
The model SEDs are plotted for multiple model apertures of 10000 AU (black line), 
and 1000 AU (gray line). The model is somewhat deficient in flux at long wavelengths; however, this depends
on both the level of external heating and the large-scale density profile and envelope structure.
 The IRS spectrum has a 3\farcs6 slit at wavelengths below 14 \micron; at greater than 
14 \micron\ the slit size is 11\arcsec\ wide, but aperture size does not matter in this range of wavelength since
the emission is all from very small radii.}
\label{modelseds}
\end{figure}

\begin{figure}
\begin{center}
\includegraphics[scale=0.5]{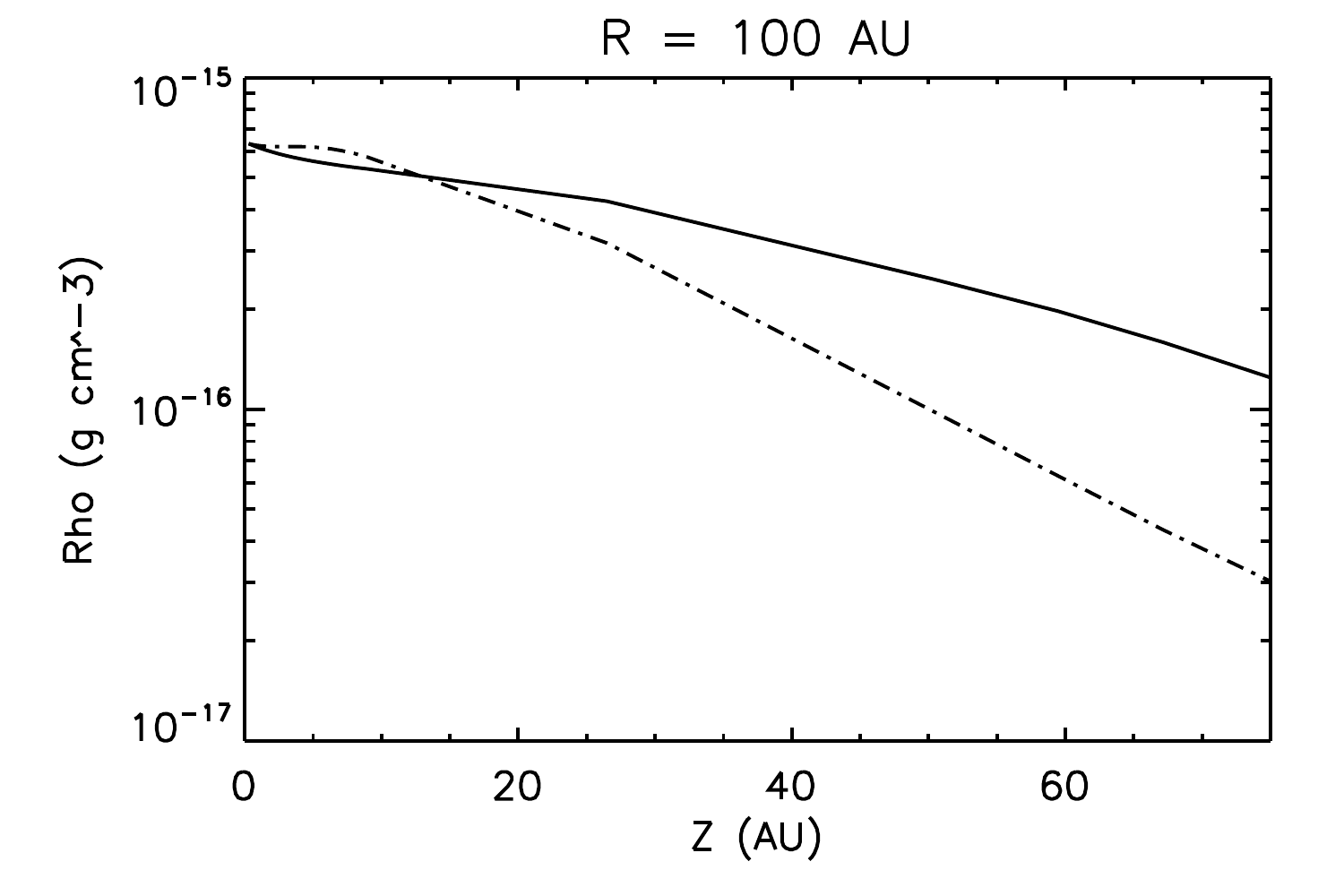}
\includegraphics[scale=0.5]{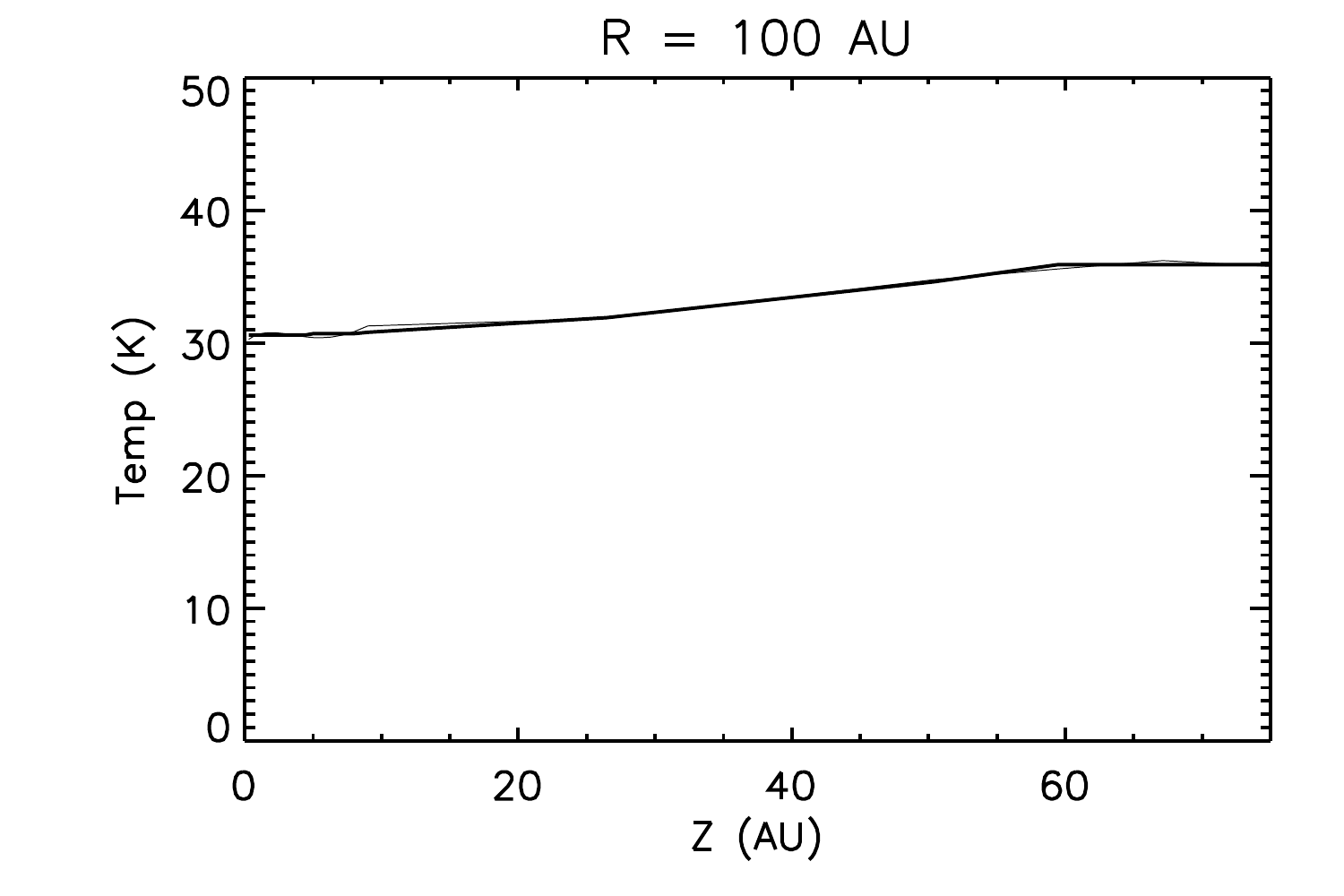}
\end{center}
\caption{
Plot of disk density (left) and dust temperature (right) at $R$ = 100 AU as a 
function of distance above the midplane (Z) for the best fitting model.
A hydrostatic equilibrium (HSEQ) solution is also plotted for the same central density and the temperatures 
shown in the right panel (dot-dashed line). The best fitting disk model disk for L1527 is more extended
than HSEQ given the input temperature profile. The midplane temperature is quite warm compared to 
Class II disks, the 30 K central temperature implies that
CO will not be frozen out and molecules should be present in the gas.
}
\label{diskstructure}
\end{figure}

The visibilities of the best fitting models
are overlaid on the data in Figure \ref{modelvis}. In Figure \ref{modelimage}, we show the model images at 
870 \micron\, and 3.4 mm and the residuals when subtracted from the data. The model images
reproduce the data quite well. Figure \ref{modelLimages} shows the models compared to 
the observed L\arcmin\ image; both models reproduce the locations of the brightest parts of scattered 
light well, but the more diffuse emission not reproduced by the models could be extended, upper layers of the disk or
scattering in the outflow cavity. The SEDs of the best fitting model are shown in Figure \ref{modelseds}; there
is some deviation at wavelengths longer than 20 \micron, but this may be due to disk vertical structure effects. We show the 
vertical density and temperatures profile at R = 100 AU for the model in Figure \ref{diskstructure}
 as well as a hydrostatic equilibrium (HSEQ) calculation based on the equatorial density and vertical
temperature distribution. This shows that the best fitting disk model has a vertical density
profile about a factor of two larger vertical heights
than HSEQ. A caveat here is that HSEQ density profile is calculated for the temperature profile
of the best fit model, given that the density structure is different in HSEQ, the temperature profile will
not be the same. Nonetheless, this serves as a simple demonstration of the disk model being more 
extended than HSEQ might allow. The parameters of the best fitting model is given in Table 5.

\subsection{Comparison with Other Observations and Models}

\subsubsection{L1527}
Previously, the highest resolution sub/millimeter data taken for L1527 were from
the Plateau de Bure Interferometer (PdBI) at 1.3 mm \citep{maury2010}. 
The CARMA and SMA data have modestly better resolution, but we were able 
to clearly resolve the disk and their observations did not. This was
due to the elongation of the PdBI synthesized beam in the north-south direction, the same
direction as the disk long axis. 
\citet{jorgensen2009} used the SMA compact array data to infer the 
disk mass by assuming that the flux at 50 k$\lambda$ was primarily from the disk. 
The disk mass of L1527 was found to be 0.029 $M_{\sun}$ (assumed T = 30 K); when adjusted
for our assumed value of $\kappa_{850\mu m}$, $M_{disk}$ = 0.015 $M_{\sun}$, larger but not inconsistent with our modeled value.
There were also VLA 7mm observations of L1527 in A configuration that detected
a small-scale disk with an apparent radius of 20 AU \citep{loinard2002}, but with substantial spatial filtering.
EVLA studies of L1527 are also underway with initial results
from D configuration were presented by \citet{melis2011}.

\subsubsection{Class 0 Sources}
Other than \citet{jorgensen2007,jorgensen2009} only a few studies have had the resolution to examine disks in Class 0 sources. \citet{brown2000}
used the JCMT-CSO interferometer to constrain disk radii in a few Class 0 sources to be $\sim$300 AU. \citet{looney2000}
surveyed a range of young stellar objects (Class 0 - II), finding evidence of a resolved disk around NGC 1333 IRAS 2A
as well as some Class I sources. \citet{chiang2012} observed L1157
with high-resolution using CARMA and do not find a resolved disk on 90 AU scales, while \citet{lee2009} found
indications of a disk in HH211 and a companion.
\citet{maury2010} published a high-resolution sample of 5 protostars observed with the
PdBI (L1527 included), none of which showed strong signs of resolved disk emission. 
A large ($R$ = 300 AU), massive ($M_{disk}$ $\sim$ 1 $M_{\sun}$)
disk was found around Serpens FIRS1 by \citet{enoch2009}; however, this system is likely
intermediate mass with the revised/increased distance estimate toward Serpens.
Finally, \citet{harvey2003} examined the B335 system, inferring that there was a 0.004 $M_{\sun}$ disk from
modeling the flattening visibility amplitudes, but they did not resolve the disk and only 
inferred a radius $<$ 100 AU. Thus, L1527 is currently in a class by itself when it comes to resolved disks
in the Class 0 phase.

 \subsubsection{Class I \& II sources}
Several disks have been imaged in the infrared and submillimeter around Class I 
protostars with disk radii between 50 and 300 AU \citep[e.g.][]{stark2006, padgett1999}.
High-resolution dust continuum observations have been taken toward IRAS 04302+2247,
finding $M_{disk}$ $\sim$ 0.07 $M_{\sun}$ and $R_{disk}$ = 300AU \citep{wolf2008}. CB26 was also resolved
in the near-infrared and millimeter by \citet{launhardt2001}, subsequent modeling by \citep{sauter2009} finds
$M_{disk}$ $\sim$ 0.3 $M_{\sun}$ and $R_{disk}$ = 200 AU. 
The characteristic radii of Class II disks peaks at about 50 AU \citep{andrews2010},
but extends to 100s of AU. These results indicate that the disk in L1527
 already has a characteristic radius similar to Class I and II sources. 
The disk mass of 0.0075 $M_{\sun}$ for L1527 is quite comparable
to what is found for Class I or II sources, but this is a poorly constrained quantity. The radial
density profile of L1527 may also be steeper than the typical $\Sigma$ $\propto$ $R^{-1}$ found in
Class II disks, but this quantity is not well-constrained since the disk is viewed edge-on.
Exact parameters of the best fit model are compared to IRAS 04302+2247 and CB26 in Table 7.

The difference in observed scale-height between L1527 and Class I/II sources is the most striking.
The models of Class I and II  sources indicate typical values of $H_{100}$ to be about 10 - 15 AU 
\citep{sauter2009,wolf2008, andrews2010}, while L1527 has $H_{100}$ = 48 AU. 
However, under the assumption of Gaussian vertical structure, L1527 is about a factor of two more
vertically extended than HSEQ, as shown in Figure \ref{diskstructure}.
However, in more evolved systems the scale heights may appear smaller due to dust settling.
In the case of HL Tau, the disk was found to be 
1.5$\times$ more vertically extended than HSEQ with H$_{100}$ = 19 AU \citep{kwon2011}. The vertical height of
the disk in L1527 appears to be substantially larger than in Class II disks.

\section{Discussion}

We have conducted radiative transfer modeling of the youngest protostellar system
know to harbor a rotationally-supported disk
in both sub/millimeter dust emission and near-infrared scattered light. 
The system was modeled with a disk surrounded by an infalling envelope with masses and radii
typical of Class II disks \citep[e.g.][]{andrews2007,andrews2009,andrews2010},
while the vertical height is substantially larger. The parameters of disks 
in these early stages represent the initial conditions of the Class I/II objects and indicate
 how likely a system might be to fragment gravitationally forming a binary system and/or giant planets. 
The principal difficultly in characterizing young disks thus far has been the surrounding envelope; at most
size-scales, the emission is a combination of disk and envelope. Thus, in order to be certain that the disk
is being detected, a system must be observed with sufficient resolution to resolve it and/or kinematic data
must find rotationally supported motion. In the case of L1527, we have both resolved the disk in the
dust continuum at multiple wavelengths and kinematic data show a rotation signature consistent
with Keplerian rotation \citep{tobin2012}.

\subsection{Evolutionary State of L1527}

Whether or not L1527 is a Class 0 or Class I protostar is an important distinction
for this work as well as the implications for disk formation theory.
It was previously discussed in \citet{tobin2008} that L1527 would probably be classified as a Class I object if it were not viewed
edge-on. This is because the submillimeter luminosity would no longer meet the criteria of
being $\ge$ 0.5\% of the bolometric luminosity \citep{andre1993}, with a similar effect on the bolometric
temperature. The classification of protostars as Class 0 based on bolometric 
temperature or submillimeter luminosity may be unreliable since all protostars seem to have
outflow cavities and these diagnostics depend on inclination \citep{jorgensen2009}.
Moreover, the classification of a protostar as Class I is not exclusive of Class 0, since the
Class I - III scheme is based on near to mid-infrared spectral indicies and Class 0 is based on submillimeter flux 
or bolometric temperature. \textit{Spitzer}
detects many well-known Class 0 protostars between 3.6 and 8\micron\ with rising SED slopes out to 
24/70 \micron\ \citep{tobin2007,jorgensen2006,seale2008,enoch2009,tobin2010a}
enabling them to be simultaneously classified as Class I and Class 0. 
Semantics aside, the main issue is whether or not L1527 is consistent with being a very young protostar or if 
it has evolved into a later phase.

Bonafide Class I sources in Taurus (e.g. IRAS 04302+2247) appear much more compact at IRAC wavelengths, and
do not have the resolved outflow cavities extending out to $\sim$10000 AU as seen in L1527 \citep{hartmann2005}. Furthermore,
the envelopes of Class I protostars also have much more compact millimeter emission \citep[i.e. L1489 IRS,][]{motte2001}; 
L1527 has a large envelope resolved in the submillimeter \citep{chandler2000}. On the other hand, 
L1527 differs from the more typical Class 0 objects that have more well-collimated outflows 
and do not have as intense scattered light nebulosities; \citet{tobin2010a} shows
examples of this. Independent of the photometric and image properties, \citet{emprechtinger2009} 
showed that younger Class 0 sources typically have higher ratios of \ntdp\ to \nthp\ than 
more evolved Class 0 or Class I sources. Toward the central protostar of L1527, 
there was not a detectable level of \ntdp, but there is \ntdp\ at larger radii \citep{tobin2013}.

The above discussion demonstrates that the observational Class of a particular system does not necessarily 
indicate the evolutionary state. \citet{rob2006} attempted to alleviate this distinction by defining ``stages''; 
however, these stages are based on model dependent quantities. Within this framework, however, 
L1527 is consistent with being a Stage 0/I protostar. Moreover, \citet{andre1993} suggested that the 
observational classification of Class 0 would mean $M_{env}$ $>$ $M_{*}$ and 
the envelope mass of L1527 exceeds the protostellar mass by about factor of five
\citep{shirley2000,chandler2000,tobin2010a}. While L1527 does not 
conform to all the general properties of a Class 0 or Class I source,
it is clearly in an early evolutionary state, but likely not among the youngest of Class 0 protostars.

\subsection{Disk Structure}

The radiative transfer modeling enabled us to derive likely values for structural parameters of the 
protostellar disk, within the limitations of our parametrized disk model. We find that a
disk radius of 125 AU provides the best fit to the data. This is not terribly different from
the 190 AU radius indicated in Paper I, but this value is likely more robust given the additional observational 
constraints. Moreover, this radius should not be thought of as a hard outer edge of the disk, and perhaps more as
a transition point between and envelope-dominated density/kinematic structure and that of the disk. A more physical treatment
of the disk-envelope interface may be necessary in future modeling 
to better understand the assembly of the proto-planetary disk.

The mass of the disk in L1527 is not large, only 0.0075 $M_{\sun}$, consistent with the optically
thin calculation. However, this value should be regarded
with extreme caution, give the uncertainties in $\beta$, assumed dust opacity law, 
and the dust to gas ratio \citep[e.g.][]{ricci2010}. 
While the dust opacity directly affects our measurement of the mass, it should 
not strongly affect the other properties of our model given that most of the emission 
is optically thin and we expect that $\kappa_{850\mu m}$ should be \textit{lower} than our assumed value.
We calculated that emission from our disk model, integrated along the line of sight, will become 
optically thick ($\tau$ = 1) at projected radial distances of 11.5 AU and 9 AU at 870 \micron\ and 3.4 mm respectively;
much smaller than our spatial resolution. The enclosed masses within 11.5 AU and 9 AU radii are 
7.4 $\times$ 10$^{-4}$ $M_{\sun}$ and 6.1 $\times$ 10$^{-4}$ $M_{\sun}$, consistent with our simplistic
mass measurements in Section 3.5 being smaller than the total mass of the model.
Furthermore, there could be a dead zone at small radii that is optically thick, harboring a substantial
amount of hidden mass and we would not know about it with the current observations. The brightness temperatures
of the disk have a maximum value of T$_B$ = 12 K at 870 \micron, while the compact disk structure
observed at 7 mm by \citep{loinard2002} has T$_B$ = 50 K indicating that it may be optically thick at 7 mm.

A surprising result from our study is the large scale-height of the disk. The disk scale height $H$ is proportional to
$R^{1.3}$, with $H_{100}$ = 48 AU. For comparison, the 
analytic solution for an irradiated disk is $H$ $\propto$ $R^{1.25}$ and $R^{1.125}$ for
a steady, viscous accretion disk. Figure 10 shows the vertical density and 
temperature profiles of the disk in L1527. For comparison, the vertical hydrostatic
equilibrium solution is also plotted assuming the same central density and dust temperature structure from the model.
Using the dynamical mass measurement of 0.19 $M_{\sun}$, we find that the disk model is about a factor of two more extended
than vertical HSEQ.  Thus, even without the physics of the envelope infalling onto
the disk, our model is able to reproduce the observations quite well; infall of the envelope
onto the disk could make it more vertically extended than HSEQ.

The implied surface density profile is $\Sigma$ $\propto$ $R^{-1.2}$, inferred 
from $\rho$ $\propto$ $R^{-2.5}$ and $H$ $\propto$
$R^{1.3}$. While this parameter is not well-constrained, 
it is slightly steeper than the typical $\Sigma$ $\propto$ $R^{-1}$. This is expected
for the more evolved systems because angular momentum transport from the 
mass accretion process will redistribute the mass from its initial configuration. We do
caution that the best fit model has the minimum possible surface density distribution in the model grid.
The implied surface density distribution by infall from a TSC envelope without 
mass redistribution is $\Sigma$ $\propto$ $R^{-1.5}$ \citep{hartmann2009}; in 
agreement with one of our model cases.
Our modeled surface density profile meets the criteria for gravitational instability 
of Toomre Q = 1 ($\Sigma$ $\propto$ $R^{-1.75}$), assuming
T $\propto$ $R^{-3/4}$. However, if the disk mass is accurate, the 
protostar is more than order of magnitude more massive than the disk. $M_{disk}$ may be closer to $M_{*}$ in reality as
the dust opacities may be over estimated and there could be optically thick dust on scales smaller than our beam.
We caution that not too many conclusions can be drawn from the surface density profile because we are not directly sensitive
to the surface density and must infer it from the radial and vertical structure.

\subsection{Caveats}

The best-fitting models of the disk-envelope system in L1527 are inherently non-unique. The models being employed represent the most
idealized form of protostellar envelope and disk structures. Robust statistical measures can only be
reliably used if the model is known to represent the observations \textit{a priori}; therefore, meaningful
confidence levels cannot be assigned to the model parameters. Given that we are modeling
a protostellar system with known non-axisymmetries \citep{tobin2010a,tobin2011}, the models cannot be expected to fully reproduce 
the observations. There are also degeneracies between certain parameters because of the edge-on orientation, as we will discuss later in this section.
Moreover, some of the parameters fit may be abstractions of missing physics i.e. infall of the envelope onto the disk.

For instance, the radial density profile we find in the L1527 disk could be an artifact 
of our assumption of constant dust opacity at all
radii. There could be preferentially smaller dust grains at large radii than at smaller radii, resulting from the influx
of smaller grains from the infalling envelope and subsequent grain growth in the inner disk \citep{perez2012}.
However, this scenario does not initially seem likely given that the spectral index of dust emission is constant at uv-distances
$>$ 10 k$\lambda$ (Figure \ref{alpha_uvd}). One might expect that if the dust opacity were changing with radius in the disk and/or envelope,
we would expect to see the spectral index change with uv-distance. But, 10 k$\lambda$ corresponds to spatial scales of $\sim$25\arcsec;
\citet{shirley2011} measures an integrated 3 mm flux density of 33 mJy toward L1527, consistent with our visibility amplitudes in
Figure \ref{visibilities}, suggesting that the disk is emitting most flux on these spatial scales and the envelope
only becomes dominant at smaller spatial frequencies.

The degeneracies between parameters also adds uncertainty to our modeling. Some of the most 
significant degeneracies are: radial density profile to disk radius/mass/$\beta$,
 initial scale height to disk radius/flaring, and disk mass to $\beta$. However,
we are able to mitigate but not retire these degeneracies with the SED fit to the IRS spectral 
range and the L\arcmin\ image comparison. Higher resolution sub/millimeter data and fitting of the visibilities in two 
dimensions are both needed to better constrain disk parameters.

\subsection{Millimeter-wave Dust Opacity}

A particularly curious result is the low value of $\beta$, which we allowed to vary in our modeling.
This could result from both systematic and astrophysical effects. An obvious culprit could
be systematic errors in the absolute flux scale. Each dataset is calibrated as well as possible;
the observations in different array configurations agree quite well in the regions of overlap in the uv-plane, even
when different flux calibrators are used and when there were long time intervals between observations. 
Moreover, we did not find large discrepancies between our
measured calibrator fluxes and those in the SMA and CARMA calibrator databases. Nonetheless, there is uncertainty
in the planet models used for absolute flux calibration and there are differences between those used in MIRIAD for
CARMA and those in MIR for the SMA. Variability of the source is another 
possibility, but the agreement of the measured flux over long time intervals seems to rule this out. 
Moreover, the SMA and CARMA observations at the highest resolution were taken
a little over a month apart, mitigating the possibility of large flux variations.

Thus, the low value of $\beta$ seems to be real, confirming previous indications
of a shallow $\beta$ \citep{melis2011,scaife2012}. In addition,
previous single-dish observations of the envelope indicated 
$\beta$ $\sim$ 1 \citep{shirley2011}. Moreover, observations of Class II sources in Taurus
also show evidence for low values of $\beta$ \citep{andrews2005,ricci2010} as well
as Class 0 protostars \citep{kwon2009,chiang2012}; therefore, a shallow $\beta$ in
the inner envelope and disk may be a realistic expectation. 

The low value of $\beta$ can be explained by
a population of large dust grains in the disk and inner envelope, as shown by opacity curves
for large grains and different power-law size distributions \citep{hartmann2008,ricci2010}. 
The dust opacity models presented in \citet{ricci2010} have 
$\kappa_{850\mu m}$ $\sim$ 2.5 cm$^{2}$ g$^{-1}$ (dust only opacity), with $\beta$ $\sim$0.25.
Two models yield values of $\beta$ consistent with L1527, one with 
a power-law size distribution of $N(a)$ $\propto$ $a^{-2.5}$ and 
a maximum grain size of 1~cm, and another with $N(a)$ $\propto$ $a^{-3.0}$ 
and a maximum grain size of 10~cm. For comparison, we
used $\kappa_{850\mu m}$~=~3.5~cm$^{2}$~g$^{-1}$ for the sub/millimeter modeling and
$\kappa_{850\mu m}$~=~0.6~cm$^{2}$~g$^{-1}$ for the Monte Carlo model ($\beta$ $\sim$ 1.93);
\citet{ossenkopf1994} gives $\kappa_{850\mu m}$~=~1.82~cm$^{2}$~g$^{-1}$ 
($\beta$ $\sim$ 1.75). Thus, the \citet{ricci2010} model seems attractive 
for the sub/millimeter regime and could adjust the
disk mass upward by 1.6$\times$.

While, these opacity models for large dust grains are consistent with our
observations, could such large dust grains have formed in the relatively short
life of a Class 0 protostellar disk? Models of dust grain evolution considering
the formation and growth of the disk, with infall from the envelope have been 
computed by \citet{birnstiel2010}. These models showed that grains can grow 
rapidly from micron-sizes to cm-sizes within $\sim$50 kyr, out to radii of $\sim$60 AU.
Therefore, it seems possible that grain growth to cm-sizes could explain the
shallow $\beta$. In addition, there is evidence
that the dust grains in the envelopes are at least micron-sized \citep{steinacker2010}
and possibly larger \citep{shirley2011,kwon2009,chiang2012}, so the disks may be seeded
with larger grains to start with. 

We note that any large dust grains
must be concentrated in the disk, because 
larger grains throughout the envelope would
broaden the 10~\micron\ silicate absorption feature and observations show this
feature to be quite sharp in Figure \ref{modelseds}. This makes sense
because grains will grow fastest where the densities are largest, i.e. the disk.
Such large grains must also be settled to the disk midplane \citep{dullemond2004,dalessio2006}, because large dust
grains throughout the disk would also cause the mid-infrared SED 
to from 10~\micron\ to 70~\micron\ to not fit the data \citep{tobin2010b}. Furthermore,
the larger the grains grow, the more rapidly they settle to the disk midplane;
the settling timescale is shorter than the grain growth timescale \citep{dullemond2004}.
Therefore, we suggest that a population of large dust grains, responsible for 
much of the emission at sub/millimeter wavelengths and shallow opacity
spectral index, are settled in the disk midplane 
with a smaller scale height than the disk observed in scattered light.
Nevertheless, it remains possible that much of the millimeter-wave emission 
is coming from an optically thick, dense disk midplane, similar to the 
model by \citep{grafe2013} for IRAS 04302+2247. In this scenario, the upper layers of the
disk, responsible for the L\arcmin\ scattered light, are tracing a different size distribution of
dust grains. However, recall that the millimeter-wave SED is consistent with $\beta$ $\sim$ 0 
out to 7 mm, where we would certainly expect the emission to be optically thin.

\subsection{Constraints on Disk Formation and Evolution}

If L1527 reflects the state of disks toward the end of the Class 0 phase, then we can infer
that disks must grow rapidly within the Class 0 phase, as it only lasts $\sim$ 0.5 Myr \citep{evans2009}.
Idealized magnetic collapse models predict that disks no greater than 10 AU 
would be observed in Class 0 systems, within 4 $\times$ 10$^4$ yr of collapse \citet{dapp2010}.
 A radius of 125 AU may be a bit larger than 
expected for such a young source \citep{dapp2012}; however, we should expect to find a variety of disk
radii in young protostellar systems, since star forming cores will have a variety of 
angular momenta \citep{goodman1993, caselli2002, chen2007,tobin2011}.
Therefore, the models of \citet{dapp2012} either need to start with more angular momentum or magnetic braking
is not as effective at retarding the formation large disks as their models suggest.
On the other hand, \citet{vorobyov2010} showed that disks similar in size to
L1527 can form before the end of the Class 0 phase with only modest initial angular 
velocities, but this simulation ignores the effects of magnetic fields. \citet{joos2012}
then finds that large disks can form in ideal MHD simulations, but only if the initial magnetic field
direction is misaligned with the outflow, a phenomenon that has been recently observed
in a number of protostars, L1527 included \citep{hull2013}.

\citet{yorke1999} also conducted simulations of disk formation in two dimensions, but 
in an edge-on axisymmetric geometry, also ignoring magnetic fields. Large disks also 
formed in these simulations; however, the disks (the low-mass Case H and J)
 have $H_{100}$ = 8 AU (extrapolated from given values at $R$ = 500, 1000 AU), substantially smaller than
our models for L1527. The disks are highly flared ($H$ $\propto$ $R^{1.34}$), while the initial scale heights are 
quite small. Note that these values are from the end of the 
simulation, where the disk has grown to $R$ $>$ 1500 AU and the amount of mass in the sink cell is 0.6 $M_{\sun}$.
The infall onto the disk does generate a shock, but this shock is $\sim$ 8 scale heights above the midplane, creating 
a $\sim$40 AU region of relatively constant density in the z direction. This shocked region
could qualitatively explain the diffuse L\arcmin-band scattered light extended so high above the 
disk midplane, in addition to non-Gaussian vertical structure.

Overall, our results suggest that by the end of the Class 0 phase, it is possible for disks to have 
grown to a substantial fraction of their ultimate radius, as evidenced
by Class I and II disks not being substantially larger than the disk in L1527. This is in agreement with hydrodynamic simulations
and could be consistent with simulations that include magnetic fields, depending on initial conditions.
The disk also must become substantially less flared and vertically extended between the Class 0 to Class I/II phases.
This process likely involves both dust settling \citep{dalessio2006} and contraction toward 
the midplane as the stellar mass grows to maintain HSEQ. 

The observation of a shallow dust opacity spectral index in the disk of L1527, taken
with the observations of $\beta$~$<$~1 in other protostars \citep{kwon2009,chiang2012} seems
to indicate that grain growth may proceed rapidly during the protostellar phase. Thus, disks
in the Class I/II phase may have already have large grains, perhaps enabling planetesimal growth
at in the early in the star formation process.

\subsection{Timescales for Disk and Protostellar Mass Growth}

Given that, we have a constraint on the protostellar mass, the disk 
radius, and the large-scale velocity gradient from \citet{tobin2011},
we can compare these values to the \citet{tsc1984} and \citet{shu1977} 
timescales for disk and protostellar growth. While these relationships
are simplistic and ignore magnetic fields, they are illustrative of the
timescales involved for disk assembly and mass accumulation. Following 
\citet{hartmann2009}, $R_C$ grows as 
\begin{equation}
R_C(t) = \frac{m_{\circ}^3}{16} c_s t^3 \Omega_{\circ}^2
\end{equation}
where $m_{\circ}$ = 0.975, $c_s$ = 0.2 \kms\ and $\Omega_{\circ}$ = 2.2 \kmspc\ \citep{tobin2011}.
The protostellar mass then grows as
\begin{equation}
M_* = \frac{m_{\circ}c_s^3t}{G}
\end{equation}
where G is the gravitational constant. Application of these equations indicates that 
it would take 1.1 $\times$ 10$^5$ yr for the protostar 
to accumulate 0.19 $M_{\sun}$ and 2.16 $\times$ 10$^5$ yr to
grow the disk to 125 AU, a factor of $\sim$2 difference.
Thus, in the time it took to build the disk to 125 AU, the protostellar mass 
should be 0.4 $M_{\sun}$. This could mean that the additional mass is being built up
in the disk, but it would have to be optically thick and on small spatial scales,
below our resolution limit. Furthermore, the disk timescale hinges greatly on 
the assumed initial cloud rotation. Both \citet{ohashi1997} \& \citet{tobin2011}
show that the outer envelope velocity gradient is in the \textit{opposite} 
direction of the disk rotation. If we instead adopt the inner envelope velocity 
gradient of 12.1 \kmspc, the timescale form a 125 AU disk is reduced to 7 $\times$ 10$^4$ yr, in closer
agreement to the mass accretion timescale.

\subsection{Future Prospects for Class 0 disks}
In order to clearly detect the disk in L1527, we had to observe L1527 at 
the limit of angular resolution available to current millimeter
interferometers (SMA, CARMA, and Cycle 0 ALMA). The disk in L1527 was not clearly resolved until it 
was observed with a spatial resolution roughly 5x finer than its modeled diameter.
Therefore, we suggest that in order to detect disks embedded in dense envelopes,
they must be over-resolved by a factor of about five compared to the expected diameter, in 
order to distinguish the disk from the envelope. 
Thus, with the current sub/millimeter interferometers, we should not expect to clearly resolve 
disks around Class 0 protostars in star forming regions more distant than Taurus or Perseus; even then
only the largest Class 0 disks will be able to be resolved. 
When ALMA is fully online this landscape will change dramatically. 
Disks as small as 30 AU around the youngest protostars in Taurus should be resolved and 60 AU in Orion; 
the Jansky VLA also has this capability of observations at 7~mm. 
Therefore, the future prospects of resolving disks around Class 0 protostars
may be much brighter than suggested by \citet{maury2010} and \citet{dapp2012}. Moreover,
the possibility of observe molecular line emission from the disks with ALMA should enable the masses of a large
sample of Class 0 protostellar objects to be characterized.

\section{Conclusions}
We have presented analysis and radiative transfer modeling of high-resolution observations toward
the protostar L1527 at $\lambda$ = 870 \micron\ and 3.4 mm with the SMA and CARMA. 
We found that the protostellar system can be well-modeled
with an extended disk, using the Monte Carlo radiative transfer 
codes of \citet{whitney2003} and \citet{brinch2010}. 

1. We modeled the disk emission by fitting the one dimensional visibilities, sub/millimeter
images, multi-wavelength SED, and L\arcmin\ images. The best fitting model has 
a radius of 125 AU, mass of 0.0075 $M_{\sun}$, a radial density profiles 
proportional to $R^{-2.5}$, power-law flaring with 
$H$ $\propto$ $R^{1.3}$ (assuming Gaussian vertical structure), vertical scale heights at 100 AU of 48 AU, and a dust
opacity spectral index of $\beta$ = 0.25. The most robust parameters are the disk radii and vertical height
as these are well-constrained by the resolved image data in the sub/millimeter and L\arcmin\
scattered light. The disk vertical structure is not drastically out of hydrostatic equilibrium, a factor of two
larger, but this is subject to the assumptions of the Gaussian vertical density profile. 
The most degenerate parameters are the disk mass, radial density profile. The disk mass 
is dependent on the assumed dust opacity, the radial structure is not well enough resolved.

2. The millimeter spectral index from the integrated fluxes at 870 \micron, 3.4 mm, and other 
work suggest a spectral slope in the millimeter
of $S_{\lambda}$ $\propto$ $\lambda^{-2}$. Assuming optically thin and isothermal emission, 
this suggests that the dust opacity spectral index $\beta$ = 0. We then examined the spectral 
index from 870 \micron\ to 3.4 mm as a function of uv-distance, 
finding that at all spatial scales $\beta$ is consistent with zero. The modeled
dust opacity spectral index of 0.25 is consistent with the shallow spectral index. This is
the shallowest known spectral index for a Class 0 protostar and bears similarity to
what has been observed for more evolved Class II sources.
This result suggests that the dust opacities may have a shallow power-law size
distribution and maximum grain sizes of 1 - 10 cm. Models of grain growth show that
rapid growth to cm sizes is possible within the timescale of the Class 0 phase; thus, Class I/II disks
may already start with large dust grains.

3. L1527 appears most consistent with a late Class 0 source, but shares 
some characteristics of both Class I and Class 0 protostars.
Comparing the disk properties to Class I/II sources, we find L1527 to have a smaller radius and less flared, possibly
indicating an evolutionary trend of sustained disk growth and settling into the Class I phase.

{\it Facilities:} \facility{Gemini:Gillett (NIRI)}, \facility{VLA}, \facility{CARMA}, \facility{SMA}

The authors wish to thank the anonymous referee for comments which improved the manuscript.
J. T. acknowledges support provided by NASA through Hubble Fellowship 
grant \#HST-HF-51300.01-A awarded by the Space Telescope Science Institute, which is 
operated by the Association of Universities for Research in Astronomy, 
Inc., for NASA, under contract NAS 5-26555. L. H. and J. T. acknowledge partial
support from the University of Michigan. H.-F. C. acknowledges
support from the National Aeronautics and Space
Administration through the NASA Astrobiology Institute under
Cooperative Agreement No. NNA09DA77A issued through the Office of
Space Science. L.W.L. and H.-F. C. acknowledge support from the Laboratory for Astronomical 
Imaging at the University of Illinois and the NSF under grant AST-07-09206.
P. D. acknowledges a grant from PAPIIT-UNAM.
L. L. acknowledges the support of DGAPA, UNAM, CONACyT (M\'exico), 
and the Alexander von Humboldt Stiftung for financial support.
Support for CARMA construction was derived from the states of Illinois, California, and Maryland, 
the James S. McDonnell Foundation, the Gordon and Betty Moore Foundation, the Kenneth T. and 
Eileen L. Norris Foundation, the University of Chicago, the Associates of the California 
Institute of Technology, and the National Science Foundation. Ongoing CARMA development 
and operations are supported by the National Science Foundation under a cooperative 
agreement, and by the CARMA partner universities. The Submillimeter Array is a joint 
project between the Smithsonian Astrophysical Observatory and the Academia Sinica 
Institute of Astronomy and Astrophysics and is funded by the Smithsonian 
Institution and the Academia Sinica. The National Radio Astronomy 
Observatory is a facility of the National Science Foundation 
operated under cooperative agreement by Associated Universities, Inc.

\begin{small}
\bibliographystyle{apj}
\bibliography{ms}

\begin{thebibliography}{94}
\expandafter\ifx\csname natexlab\endcsname\relax\def\natexlab#1{#1}\fi

\bibitem[{{Allen} {et~al.}(2003){Allen}, {Li}, \& {Shu}}]{allen2003}
{Allen}, A., {Li}, Z., \& {Shu}, F.~H. 2003, \apj, 599, 363

\bibitem[{{Andre} {et~al.}(1993){Andre}, {Ward-Thompson}, \&
  {Barsony}}]{andre1993}
{Andre}, P., {Ward-Thompson}, D., \& {Barsony}, M. 1993, \apj, 406, 122

\bibitem[{{Andrews} \& {Williams}(2005)}]{andrews2005}
{Andrews}, S.~M., \& {Williams}, J.~P. 2005, \apj, 631, 1134

\bibitem[{{Andrews} \& {Williams}(2007)}]{andrews2007}
---. 2007, \apj, 659, 705

\bibitem[{{Andrews} {et~al.}(2009){Andrews}, {Wilner}, {Hughes}, {Qi}, \&
  {Dullemond}}]{andrews2009}
{Andrews}, S.~M., {Wilner}, D.~J., {Hughes}, A.~M., {Qi}, C., \& {Dullemond},
  C.~P. 2009, \apj, 700, 1502

\bibitem[{{Andrews} {et~al.}(2010){Andrews}, {Wilner}, {Hughes}, {Qi}, \&
  {Dullemond}}]{andrews2010}
---. 2010, \apj, 723, 1241

\bibitem[{{Birnstiel} {et~al.}(2010){Birnstiel}, {Dullemond}, \&
  {Brauer}}]{birnstiel2010}
{Birnstiel}, T., {Dullemond}, C.~P., \& {Brauer}, F. 2010, \aap, 513, A79

\bibitem[{{Brinch} \& {Hogerheijde}(2010)}]{brinch2010}
{Brinch}, C., \& {Hogerheijde}, M.~R. 2010, \aap, 523, A25

\bibitem[{{Brown} {et~al.}(2000){Brown}, {Chandler}, {Carlstrom}, {Hills},
  {Lay}, {Matthews}, {Richer}, \& {Wilson}}]{brown2000}
{Brown}, D.~W., {Chandler}, C.~J., {Carlstrom}, J.~E., {Hills}, R.~E., {Lay},
  O.~P., {Matthews}, B.~C., {Richer}, J.~S., \& {Wilson}, C.~D. 2000, \mnras,
  319, 154

\bibitem[{{Calvet} {et~al.}(2005){Calvet}, {D'Alessio}, {Watson},
  {Franco-Hern{\'a}ndez}, {Furlan}, {Green}, {Sutter}, {Forrest}, {Hartmann},
  {Uchida}, {Keller}, {Sargent}, {Najita}, {Herter}, {Barry}, \&
  {Hall}}]{calvet2005}
{Calvet}, N., {D'Alessio}, P., {Watson}, D.~M., {Franco-Hern{\'a}ndez}, R.,
  {Furlan}, E., {Green}, J., {Sutter}, P.~M., {Forrest}, W.~J., {Hartmann}, L.,
  {Uchida}, K.~I., {Keller}, L.~D., {Sargent}, B., {Najita}, J., {Herter},
  T.~L., {Barry}, D.~J., \& {Hall}, P. 2005, \apjl, 630, L185

\bibitem[{{Caselli} {et~al.}(2002){Caselli}, {Benson}, {Myers}, \&
  {Tafalla}}]{caselli2002}
{Caselli}, P., {Benson}, P.~J., {Myers}, P.~C., \& {Tafalla}, M. 2002, \apj,
  572, 238

\bibitem[{{Cassen} \& {Moosman}(1981)}]{cassen1981}
{Cassen}, P., \& {Moosman}, A. 1981, \icarus, 48, 353

\bibitem[{{Chandler} \& {Richer}(2000)}]{chandler2000}
{Chandler}, C.~J., \& {Richer}, J.~S. 2000, \apj, 530, 851

\bibitem[{{Chen} {et~al.}(2007){Chen}, {Launhardt}, \& {Henning}}]{chen2007}
{Chen}, X., {Launhardt}, R., \& {Henning}, T. 2007, \apj, 669, 1058

\bibitem[{{Chiang} {et~al.}(2008){Chiang}, {Looney}, {Tassis}, {Mundy}, \&
  {Mouschovias}}]{chiang2008}
{Chiang}, H., {Looney}, L.~W., {Tassis}, K., {Mundy}, L.~G., \& {Mouschovias},
  T.~C. 2008, \apj, 680, 474

\bibitem[{{Chiang} {et~al.}(2012){Chiang}, {Looney}, \& {Tobin}}]{chiang2012}
{Chiang}, H., {Looney}, L.~W., \& {Tobin}, J.~J. 2012, \apj, 709, 470

\bibitem[{{D'Alessio} {et~al.}(2001){D'Alessio}, {Calvet}, \&
  {Hartmann}}]{dalessio2001}
{D'Alessio}, P., {Calvet}, N., \& {Hartmann}, L. 2001, \apj, 553, 321

\bibitem[{{D'Alessio} {et~al.}(2006){D'Alessio}, {Calvet}, {Hartmann},
  {Franco-Hern{\'a}ndez}, \& {Serv{\'{\i}}n}}]{dalessio2006}
{D'Alessio}, P., {Calvet}, N., {Hartmann}, L., {Franco-Hern{\'a}ndez}, R., \&
  {Serv{\'{\i}}n}, H. 2006, \apj, 638, 314

\bibitem[{{Dapp} \& {Basu}(2010)}]{dapp2010}
{Dapp}, W.~B., \& {Basu}, S. 2010, \aap, 521, L56+

\bibitem[{{Dapp} {et~al.}(2011){Dapp}, {Basu}, \& {Kunz}}]{dapp2012}
{Dapp}, W.~B., {Basu}, S., \& {Kunz}, M.~W. 2011, ArXiv e-prints

\bibitem[{{Dullemond} \& {Dominik}(2004)}]{dullemond2004}
{Dullemond}, C.~P., \& {Dominik}, C. 2004, \aap, 421, 1075

\bibitem[{{Dutrey} {et~al.}(1996){Dutrey}, {Guilloteau}, {Duvert}, {Prato},
  {Simon}, {Schuster}, \& {Menard}}]{dutrey1996}
{Dutrey}, A., {Guilloteau}, S., {Duvert}, G., {Prato}, L., {Simon}, M.,
  {Schuster}, K., \& {Menard}, F. 1996, \aap, 309, 493

\bibitem[{{Eisner}(2012)}]{eisner2012}
{Eisner}, J.~A. 2012, \apj, 755, 23

\bibitem[{{Emprechtinger} {et~al.}(2009){Emprechtinger}, {Caselli}, {Volgenau},
  {Stutzki}, \& {Wiedner}}]{emprechtinger2009}
{Emprechtinger}, M., {Caselli}, P., {Volgenau}, N.~H., {Stutzki}, J., \&
  {Wiedner}, M.~C. 2009, \aap, 493, 89

\bibitem[{{Enoch} {et~al.}(2009){Enoch}, {Evans}, {Sargent}, \&
  {Glenn}}]{enoch2009}
{Enoch}, M.~L., {Evans}, N.~J., {Sargent}, A.~I., \& {Glenn}, J. 2009, \apj,
  692, 973

\bibitem[{{Espaillat} {et~al.}(2007){Espaillat}, {Calvet}, {D'Alessio},
  {Hern{\'a}ndez}, {Qi}, {Hartmann}, {Furlan}, \& {Watson}}]{espaillat2007}
{Espaillat}, C., {Calvet}, N., {D'Alessio}, P., {Hern{\'a}ndez}, J., {Qi}, C.,
  {Hartmann}, L., {Furlan}, E., \& {Watson}, D.~M. 2007, \apjl, 670, L135

\bibitem[{{Espaillat} {et~al.}(2008){Espaillat}, {Calvet}, {Luhman},
  {Muzerolle}, \& {D'Alessio}}]{espaillat2008}
{Espaillat}, C., {Calvet}, N., {Luhman}, K.~L., {Muzerolle}, J., \&
  {D'Alessio}, P. 2008, \apjl, 682, L125

\bibitem[{{Evans} {et~al.}(2009){Evans}, {Dunham}, {J{\o}rgensen}, {Enoch},
  {Mer{\'{\i}}n}, {van Dishoeck}, {Alcal{\'a}}, {Myers}, {Stapelfeldt},
  {Huard}, {Allen}, {Harvey}, {van Kempen}, {Blake}, {Koerner}, {Mundy},
  {Padgett}, \& {Sargent}}]{evans2009}
{Evans}, N.~J., {Dunham}, M.~M., {J{\o}rgensen}, J.~K., {Enoch}, M.~L.,
  {Mer{\'{\i}}n}, B., {van Dishoeck}, E.~F., {Alcal{\'a}}, J.~M., {Myers},
  P.~C., {Stapelfeldt}, K.~R., {Huard}, T.~L., {Allen}, L.~E., {Harvey}, P.~M.,
  {van Kempen}, T., {Blake}, G.~A., {Koerner}, D.~W., {Mundy}, L.~G.,
  {Padgett}, D.~L., \& {Sargent}, A.~I. 2009, \apjs, 181, 321

\bibitem[{{Galli} {et~al.}(2006){Galli}, {Lizano}, {Shu}, \&
  {Allen}}]{galli2006}
{Galli}, D., {Lizano}, S., {Shu}, F.~H., \& {Allen}, A. 2006, \apj, 647, 374

\bibitem[{{Goodman} {et~al.}(1993){Goodman}, {Benson}, {Fuller}, \&
  {Myers}}]{goodman1993}
{Goodman}, A.~A., {Benson}, P.~J., {Fuller}, G.~A., \& {Myers}, P.~C. 1993,
  \apj, 406, 528

\bibitem[{{Gr{\"a}fe} {et~al.}(2013){Gr{\"a}fe}, {Wolf}, {Guilloteau},
  {Dutrey}, {Stapelfeldt}, {Pontoppidan}, \& {Sauter}}]{grafe2013}
{Gr{\"a}fe}, C., {Wolf}, S., {Guilloteau}, S., {Dutrey}, A., {Stapelfeldt}, K.,
  {Pontoppidan}, K., \& {Sauter}, J. 2013, ArXiv e-prints

\bibitem[{{Hartmann}(2008)}]{hartmann2008}
{Hartmann}, L. 2008, Physica Scripta Volume T, 130, 014012

\bibitem[{{Hartmann}(2009)}]{hartmann2009}
---. 2009, {Accretion Processes in Star Formation: Second Edition}, ed.
  {Hartmann, L.} (Cambridge University Press)

\bibitem[{{Hartmann} {et~al.}(2005){Hartmann}, {Megeath}, {Allen}, {Luhman},
  {Calvet}, {D'Alessio}, {Franco-Hernandez}, \& {Fazio}}]{hartmann2005}
{Hartmann}, L., {Megeath}, S.~T., {Allen}, L., {Luhman}, K., {Calvet}, N.,
  {D'Alessio}, P., {Franco-Hernandez}, R., \& {Fazio}, G. 2005, \apj, 629, 881

\bibitem[{{Harvey} {et~al.}(2003){Harvey}, {Wilner}, {Myers}, \&
  {Tafalla}}]{harvey2003}
{Harvey}, D.~W.~A., {Wilner}, D.~J., {Myers}, P.~C., \& {Tafalla}, M. 2003,
  \apj, 596, 383

\bibitem[{{Hennebelle} \& {Fromang}(2008)}]{hennebelle2008}
{Hennebelle}, P., \& {Fromang}, S. 2008, \aap, 477, 9

\bibitem[{{Ho} {et~al.}(2004){Ho}, {Moran}, \& {Lo}}]{ho2004}
{Ho}, P.~T.~P., {Moran}, J.~M., \& {Lo}, K.~Y. 2004, \apjl, 616, L1

\bibitem[{{Hughes} {et~al.}(2009){Hughes}, {Andrews}, {Espaillat}, {Wilner},
  {Calvet}, {D'Alessio}, {Qi}, {Williams}, \& {Hogerheijde}}]{hughes2009}
{Hughes}, A.~M., {Andrews}, S.~M., {Espaillat}, C., {Wilner}, D.~J., {Calvet},
  N., {D'Alessio}, P., {Qi}, C., {Williams}, J.~P., \& {Hogerheijde}, M.~R.
  2009, \apj, 698, 131

\bibitem[{{Hull} {et~al.}(2012){Hull}, {Plambeck}, {Bolatto}, {Bower},
  {Carpenter}, {Crutcher}, {Fiege}, {Franzmann}, {Hakobian}, {Heiles}, {Houde},
  {Hughes}, {Jameson}, {Kwon}, {Lamb}, {Looney}, {Matthews}, {Mundy}, {Pillai},
  {Pound}, {Stephens}, {Tobin}, {Vaillancourt}, {Volgenau}, \&
  {Wright}}]{hull2013}
{Hull}, C.~L.~H., {Plambeck}, R.~L., {Bolatto}, A.~D., {Bower}, G.~C.,
  {Carpenter}, J.~M., {Crutcher}, R.~M., {Fiege}, J.~D., {Franzmann}, E.,
  {Hakobian}, N.~S., {Heiles}, C., {Houde}, M., {Hughes}, A.~M., {Jameson}, K.,
  {Kwon}, W., {Lamb}, J.~W., {Looney}, L.~W., {Matthews}, B.~C., {Mundy}, L.,
  {Pillai}, T., {Pound}, M.~W., {Stephens}, I.~W., {Tobin}, J.~J.,
  {Vaillancourt}, J.~E., {Volgenau}, N.~H., \& {Wright}, M.~C.~H. 2012, ArXiv
  e-prints

\bibitem[{{Isella} {et~al.}(2009){Isella}, {Carpenter}, \&
  {Sargent}}]{isella2009}
{Isella}, A., {Carpenter}, J.~M., \& {Sargent}, A.~I. 2009, \apj, 701, 260

\bibitem[{{Joos} {et~al.}(2012){Joos}, {Hennebelle}, \& {Ciardi}}]{joos2012}
{Joos}, M., {Hennebelle}, P., \& {Ciardi}, A. 2012, \aap, 543, A128

\bibitem[{{J{\o}rgensen} {et~al.}(2007){J{\o}rgensen}, {Bourke}, {Myers}, {Di
  Francesco}, {van Dishoeck}, {Lee}, {Ohashi}, {Sch{\"o}ier}, {Takakuwa},
  {Wilner}, \& {Zhang}}]{jorgensen2007}
{J{\o}rgensen}, J.~K., {Bourke}, T.~L., {Myers}, P.~C., {Di Francesco}, J.,
  {van Dishoeck}, E.~F., {Lee}, C., {Ohashi}, N., {Sch{\"o}ier}, F.~L.,
  {Takakuwa}, S., {Wilner}, D.~J., \& {Zhang}, Q. 2007, \apj, 659, 479

\bibitem[{{J{\o}rgensen} {et~al.}(2006){J{\o}rgensen}, {Harvey}, {Evans},
  {Huard}, {Allen}, {Porras}, {Blake}, {Bourke}, {Chapman}, {Cieza}, {Koerner},
  {Lai}, {Mundy}, {Myers}, {Padgett}, {Rebull}, {Sargent}, {Spiesman},
  {Stapelfeldt}, {van Dishoeck}, {Wahhaj}, \& {Young}}]{jorgensen2006}
{J{\o}rgensen}, J.~K., {Harvey}, P.~M., {Evans}, II, N.~J., {Huard}, T.~L.,
  {Allen}, L.~E., {Porras}, A., {Blake}, G.~A., {Bourke}, T.~L., {Chapman}, N.,
  {Cieza}, L., {Koerner}, D.~W., {Lai}, S., {Mundy}, L.~G., {Myers}, P.~C.,
  {Padgett}, D.~L., {Rebull}, L., {Sargent}, A.~I., {Spiesman}, W.,
  {Stapelfeldt}, K.~R., {van Dishoeck}, E.~F., {Wahhaj}, Z., \& {Young}, K.~E.
  2006, \apj, 645, 1246

\bibitem[{{J{\o}rgensen} {et~al.}(2009){J{\o}rgensen}, {van Dishoeck},
  {Visser}, {Bourke}, {Wilner}, {Lommen}, {Hogerheijde}, \&
  {Myers}}]{jorgensen2009}
{J{\o}rgensen}, J.~K., {van Dishoeck}, E.~F., {Visser}, R., {Bourke}, T.~L.,
  {Wilner}, D.~J., {Lommen}, D., {Hogerheijde}, M.~R., \& {Myers}, P.~C. 2009,
  \aap, 507, 861

\bibitem[{{Kitamura} {et~al.}(2002){Kitamura}, {Momose}, {Yokogawa}, {Kawabe},
  {Tamura}, \& {Ida}}]{kitamura2002}
{Kitamura}, Y., {Momose}, M., {Yokogawa}, S., {Kawabe}, R., {Tamura}, M., \&
  {Ida}, S. 2002, \apj, 581, 357

\bibitem[{{Kwon} {et~al.}(2011){Kwon}, {Looney}, \& {Mundy}}]{kwon2011}
{Kwon}, W., {Looney}, L.~W., \& {Mundy}, L.~G. 2011, \apj, 741, 3

\bibitem[{{Kwon} {et~al.}(2009){Kwon}, {Looney}, {Mundy}, {Chiang}, \&
  {Kemball}}]{kwon2009}
{Kwon}, W., {Looney}, L.~W., {Mundy}, L.~G., {Chiang}, H.-F., \& {Kemball},
  A.~J. 2009, \apj, 696, 841

\bibitem[{{Lada}(1987)}]{lada1987}
{Lada}, C.~J. 1987, in IAU Symp. 115: Star Forming Regions, ed. M.~{Peimbert}
  \& J.~{Jugaku}, 1--17

\bibitem[{{Launhardt} {et~al.}(2001){Launhardt}, {Sargent}, \&
  {Zinnecker}}]{launhardt2001}
{Launhardt}, R., {Sargent}, A., \& {Zinnecker}, H. 2001, in Astronomical
  Society of the Pacific Conference Series, Vol. 235, Science with the Atacama
  Large Millimeter Array, ed. {A.~Wootten}, 134--+

\bibitem[{{Lee} {et~al.}(2009){Lee}, {Hirano}, {Palau}, {Ho}, {Bourke},
  {Zhang}, \& {Shang}}]{lee2009}
{Lee}, C., {Hirano}, N., {Palau}, A., {Ho}, P.~T.~P., {Bourke}, T.~L., {Zhang},
  Q., \& {Shang}, H. 2009, \apj, 699, 1584

\bibitem[{{Li} {et~al.}(2011){Li}, {Krasnopolsky}, \& {Shang}}]{li2011}
{Li}, Z.-Y., {Krasnopolsky}, R., \& {Shang}, H. 2011, \apj, 738, 180

\bibitem[{{Loinard} {et~al.}(2002){Loinard}, {Rodr{\'{\i}}guez}, {D'Alessio},
  {Wilner}, \& {Ho}}]{loinard2002}
{Loinard}, L., {Rodr{\'{\i}}guez}, L.~F., {D'Alessio}, P., {Wilner}, D.~J., \&
  {Ho}, P.~T.~P. 2002, \apjl, 581, L109

\bibitem[{{Looney} {et~al.}(2000){Looney}, {Mundy}, \& {Welch}}]{looney2000}
{Looney}, L.~W., {Mundy}, L.~G., \& {Welch}, W.~J. 2000, \apj, 529, 477

\bibitem[{{Looney} {et~al.}(2003){Looney}, {Mundy}, \& {Welch}}]{looney2003}
---. 2003, \apj, 592, 255

\bibitem[{{Machida} {et~al.}(2010){Machida}, {Inutsuka}, \&
  {Matsumoto}}]{machida2010}
{Machida}, M.~N., {Inutsuka}, S.-i., \& {Matsumoto}, T. 2010, \apj, 724, 1006

\bibitem[{{Maury} {et~al.}(2010){Maury}, {Andr{\'e}}, {Hennebelle}, {Motte},
  {Stamatellos}, {Bate}, {Belloche}, {Duch{\^e}ne}, \& {Whitworth}}]{maury2010}
{Maury}, A.~J., {Andr{\'e}}, P., {Hennebelle}, P., {Motte}, F., {Stamatellos},
  D., {Bate}, M., {Belloche}, A., {Duch{\^e}ne}, G., \& {Whitworth}, A. 2010,
  \aap, 512, A40+

\bibitem[{{Melis} {et~al.}(2011){Melis}, {Duch{\^e}ne}, {Chomiuk}, {Palmer},
  {Perrin}, {Maddison}, {M{\'e}nard}, {Stapelfeldt}, {Pinte}, \&
  {Duvert}}]{melis2011}
{Melis}, C., {Duch{\^e}ne}, G., {Chomiuk}, L., {Palmer}, P., {Perrin}, M.~D.,
  {Maddison}, S.~T., {M{\'e}nard}, F., {Stapelfeldt}, K., {Pinte}, C., \&
  {Duvert}, G. 2011, \apjl, 739, L7+

\bibitem[{{Mellon} \& {Li}(2008)}]{mellon2008}
{Mellon}, R.~R., \& {Li}, Z.-Y. 2008, \apj, 681, 1356

\bibitem[{{Motte} \& {Andr{\'e}}(2001)}]{motte2001}
{Motte}, F., \& {Andr{\'e}}, P. 2001, \aap, 365, 440

\bibitem[{{Ohashi} {et~al.}(1997){Ohashi}, {Hayashi}, {Ho}, \&
  {Momose}}]{ohashi1997}
{Ohashi}, N., {Hayashi}, M., {Ho}, P.~T.~P., \& {Momose}, M. 1997, \apj, 475,
  211

\bibitem[{{Ossenkopf} \& {Henning}(1994)}]{ossenkopf1994}
{Ossenkopf}, V., \& {Henning}, T. 1994, \aap, 291, 943

\bibitem[{{Padgett} {et~al.}(1999){Padgett}, {Brandner}, {Stapelfeldt},
  {Strom}, {Terebey}, \& {Koerner}}]{padgett1999}
{Padgett}, D.~L., {Brandner}, W., {Stapelfeldt}, K.~R., {Strom}, S.~E.,
  {Terebey}, S., \& {Koerner}, D. 1999, \aj, 117, 1490

\bibitem[{{Pagani} {et~al.}(2010){Pagani}, {Steinacker}, {Bacmann}, {Stutz}, \&
  {Henning}}]{steinacker2010}
{Pagani}, L., {Steinacker}, J., {Bacmann}, A., {Stutz}, A., \& {Henning}, T.
  2010, Science, 329, 1622

\bibitem[{{P{\'e}rez} {et~al.}(2012){P{\'e}rez}, {Carpenter}, {Chandler},
  {Isella}, {Andrews}, {Ricci}, {Calvet}, {Corder}, {Deller}, {Dullemond},
  {Greaves}, {Harris}, {Henning}, {Kwon}, {Lazio}, {Linz}, {Mundy}, {Sargent},
  {Storm}, {Testi}, \& {Wilner}}]{perez2012}
{P{\'e}rez}, L.~M., {Carpenter}, J.~M., {Chandler}, C.~J., {Isella}, A.,
  {Andrews}, S.~M., {Ricci}, L., {Calvet}, N., {Corder}, S.~A., {Deller},
  A.~T., {Dullemond}, C.~P., {Greaves}, J.~S., {Harris}, R.~J., {Henning}, T.,
  {Kwon}, W., {Lazio}, J., {Linz}, H., {Mundy}, L.~G., {Sargent}, A.~I.,
  {Storm}, S., {Testi}, L., \& {Wilner}, D.~J. 2012, ArXiv e-prints

\bibitem[{{P{\'e}rez} {et~al.}(2010){P{\'e}rez}, {Lamb}, {Woody}, {Carpenter},
  {Zauderer}, {Isella}, {Bock}, {Bolatto}, {Carlstrom}, {Culverhouse}, {Joy},
  {Kwon}, {Leitch}, {Marrone}, {Muchovej}, {Plambeck}, {Scott}, {Teuben}, \&
  {Wright}}]{perez2010}
{P{\'e}rez}, L.~M., {Lamb}, J.~W., {Woody}, D.~P., {Carpenter}, J.~M.,
  {Zauderer}, B.~A., {Isella}, A., {Bock}, D.~C., {Bolatto}, A.~D.,
  {Carlstrom}, J., {Culverhouse}, T.~L., {Joy}, M., {Kwon}, W., {Leitch},
  E.~M., {Marrone}, D.~P., {Muchovej}, S.~J., {Plambeck}, R.~L., {Scott},
  S.~L., {Teuben}, P.~J., \& {Wright}, M.~C.~H. 2010, \apj, 724, 493

\bibitem[{{Pi{\'e}tu} {et~al.}(2006){Pi{\'e}tu}, {Dutrey}, {Guilloteau},
  {Chapillon}, \& {Pety}}]{pietu2006}
{Pi{\'e}tu}, V., {Dutrey}, A., {Guilloteau}, S., {Chapillon}, E., \& {Pety}, J.
  2006, \aap, 460, L43

\bibitem[{{Reipurth} {et~al.}(2002){Reipurth}, {Rodr{\'{\i}}guez}, {Anglada},
  \& {Bally}}]{reipurth2002}
{Reipurth}, B., {Rodr{\'{\i}}guez}, L.~F., {Anglada}, G., \& {Bally}, J. 2002,
  \aj, 124, 1045

\bibitem[{{Ricci} {et~al.}(2010){Ricci}, {Testi}, {Natta}, {Neri}, {Cabrit}, \&
  {Herczeg}}]{ricci2010}
{Ricci}, L., {Testi}, L., {Natta}, A., {Neri}, R., {Cabrit}, S., \& {Herczeg},
  G.~J. 2010, \aap, 512, A15

\bibitem[{{Robitaille} {et~al.}(2007){Robitaille}, {Whitney}, {Indebetouw}, \&
  {Wood}}]{rob2007}
{Robitaille}, T.~P., {Whitney}, B.~A., {Indebetouw}, R., \& {Wood}, K. 2007,
  \apjs, 169, 328

\bibitem[{{Robitaille} {et~al.}(2006){Robitaille}, {Whitney}, {Indebetouw},
  {Wood}, \& {Denzmore}}]{rob2006}
{Robitaille}, T.~P., {Whitney}, B.~A., {Indebetouw}, R., {Wood}, K., \&
  {Denzmore}, P. 2006, \apjs, 167, 256

\bibitem[{{Sault} {et~al.}(1995){Sault}, {Teuben}, \& {Wright}}]{sault1995}
{Sault}, R.~J., {Teuben}, P.~J., \& {Wright}, M.~C.~H. 1995, in Astronomical
  Society of the Pacific Conference Series, Vol.~77, Astronomical Data Analysis
  Software and Systems IV, ed. {R.~A.~Shaw, H.~E.~Payne, \& J.~J.~E.~Hayes},
  433--+

\bibitem[{{Sauter} {et~al.}(2009){Sauter}, {Wolf}, {Launhardt}, {Padgett},
  {Stapelfeldt}, {Pinte}, {Duch{\^e}ne}, {M{\'e}nard}, {McCabe}, {Pontoppidan},
  {Dunham}, {Bourke}, \& {Chen}}]{sauter2009}
{Sauter}, J., {Wolf}, S., {Launhardt}, R., {Padgett}, D.~L., {Stapelfeldt},
  K.~R., {Pinte}, C., {Duch{\^e}ne}, G., {M{\'e}nard}, F., {McCabe}, C.-E.,
  {Pontoppidan}, K., {Dunham}, M., {Bourke}, T.~L., \& {Chen}, J.-H. 2009,
  \aap, 505, 1167

\bibitem[{{Scaife} {et~al.}(2012){Scaife}, {Buckle}, {Ainsworth}, {Davies},
  {Franzen}, {Grainge}, {Hobson}, {Hurley-Walker}, {Lasenby}, {Olamaie},
  {Perrott}, {Pooley}, {Ray}, {Richer}, {Rodr{\'{\i}}guez-Gonz{\'a}lvez},
  {Saunders}, {Schammel}, {Scott}, {Shimwell}, {Titterington}, \&
  {Waldram}}]{scaife2012}
{Scaife}, A.~M.~M., {Buckle}, J.~V., {Ainsworth}, R.~E., {Davies}, M.,
  {Franzen}, T.~M.~O., {Grainge}, K.~J.~B., {Hobson}, M.~P., {Hurley-Walker},
  N., {Lasenby}, A.~N., {Olamaie}, M., {Perrott}, Y.~C., {Pooley}, G.~G.,
  {Ray}, T.~P., {Richer}, J.~S., {Rodr{\'{\i}}guez-Gonz{\'a}lvez}, C.,
  {Saunders}, R.~D.~E., {Schammel}, M.~P., {Scott}, P.~F., {Shimwell}, T.,
  {Titterington}, D., \& {Waldram}, E. 2012, \mnras, 420, 3334

\bibitem[{{Seale} \& {Looney}(2008)}]{seale2008}
{Seale}, J.~P., \& {Looney}, L.~W. 2008, \apj, 675, 427

\bibitem[{{Seifried} {et~al.}(2012){Seifried}, {Banerjee}, {Pudritz}, \&
  {Klessen}}]{seifried2012}
{Seifried}, D., {Banerjee}, R., {Pudritz}, R.~E., \& {Klessen}, R.~S. 2012,
  \mnras, 423, L40

\bibitem[{{Shirley} {et~al.}(2000){Shirley}, {Evans}, {Rawlings}, \&
  {Gregersen}}]{shirley2000}
{Shirley}, Y.~L., {Evans}, II, N.~J., {Rawlings}, J.~M.~C., \& {Gregersen},
  E.~M. 2000, \apjs, 131, 249

\bibitem[{{Shirley} {et~al.}(2011){Shirley}, {Mason}, {Mangum}, {Bolin},
  {Devlin}, {Dicker}, \& {Korngut}}]{shirley2011}
{Shirley}, Y.~L., {Mason}, B.~S., {Mangum}, J.~G., {Bolin}, D.~E., {Devlin},
  M.~J., {Dicker}, S.~R., \& {Korngut}, P.~M. 2011, \aj, 141, 39

\bibitem[{{Shu}(1977)}]{shu1977}
{Shu}, F.~H. 1977, \apj, 214, 488

\bibitem[{{Stark} {et~al.}(2006){Stark}, {Whitney}, {Stassun}, \&
  {Wood}}]{stark2006}
{Stark}, D.~P., {Whitney}, B.~A., {Stassun}, K., \& {Wood}, K. 2006, \apj, 649,
  900

\bibitem[{{Takakuwa} {et~al.}(2012){Takakuwa}, {Saito}, {Lim}, {Saigo},
  {Sridharan}, \& {Patel}}]{takakuwa2012}
{Takakuwa}, S., {Saito}, M., {Lim}, J., {Saigo}, K., {Sridharan}, T.~K., \&
  {Patel}, N.~A. 2012, \apj, 754, 52

\bibitem[{{Terebey} {et~al.}(1984){Terebey}, {Shu}, \& {Cassen}}]{tsc1984}
{Terebey}, S., {Shu}, F.~H., \& {Cassen}, P. 1984, \apj, 286, 529

\bibitem[{{Tobin} {et~al.}(2013){Tobin}, {Bergin}, {Hartmann}, {Lee}, {Maret},
  {Myers}, {Looney}, {Chiang}, \& {Friesen}}]{tobin2013}
{Tobin}, J.~J., {Bergin}, E.~A., {Hartmann}, L., {Lee}, J.-E., {Maret}, S.,
  {Myers}, P.~C., {Looney}, L.~W., {Chiang}, H.-F., \& {Friesen}, R. 2013,
  \apj, 765, 18

\bibitem[{{Tobin} {et~al.}(2008){Tobin}, {Hartmann}, {Calvet}, \&
  {D'Alessio}}]{tobin2008}
{Tobin}, J.~J., {Hartmann}, L., {Calvet}, N., \& {D'Alessio}, P. 2008, \apj,
  679, 1364

\bibitem[{{Tobin} {et~al.}(2011){Tobin}, {Hartmann}, {Chiang}, {Looney},
  {Bergin}, {Chandler}, {Masqu{\'e}}, {Maret}, \& {Heitsch}}]{tobin2011}
{Tobin}, J.~J., {Hartmann}, L., {Chiang}, H.-F., {Looney}, L.~W., {Bergin},
  E.~A., {Chandler}, C.~J., {Masqu{\'e}}, J.~M., {Maret}, S., \& {Heitsch}, F.
  2011, \apj, 740, 45

\bibitem[{{Tobin} {et~al.}(2012){Tobin}, {Hartmann}, {Chiang}, {Wilner},
  {Looney}, {Loinard}, {Calvet}, \& {D'Alessio}}]{tobin2012}
{Tobin}, J.~J., {Hartmann}, L., {Chiang}, H.-F., {Wilner}, D.~J., {Looney},
  L.~W., {Loinard}, L., {Calvet}, N., \& {D'Alessio}, P. 2012, \nat, 492, 83

\bibitem[{{Tobin} {et~al.}(2010{\natexlab{a}}){Tobin}, {Hartmann}, \&
  {Loinard}}]{tobin2010b}
{Tobin}, J.~J., {Hartmann}, L., \& {Loinard}, L. 2010{\natexlab{a}}, \apjl,
  722, L12

\bibitem[{{Tobin} {et~al.}(2010{\natexlab{b}}){Tobin}, {Hartmann}, {Looney}, \&
  {Chiang}}]{tobin2010a}
{Tobin}, J.~J., {Hartmann}, L., {Looney}, L.~W., \& {Chiang}, H.
  2010{\natexlab{b}}, \apj, 712, 1010

\bibitem[{{Tobin} {et~al.}(2007){Tobin}, {Looney}, {Mundy}, {Kwon}, \&
  {Hamidouche}}]{tobin2007}
{Tobin}, J.~J., {Looney}, L.~W., {Mundy}, L.~G., {Kwon}, W., \& {Hamidouche},
  M. 2007, \apj, 659, 1404

\bibitem[{{Ulrich}(1976)}]{ulrich1976}
{Ulrich}, R.~K. 1976, \apj, 210, 377

\bibitem[{{Vorobyov}(2010)}]{vorobyov2010}
{Vorobyov}, E.~I. 2010, \apj, 723, 1294

\bibitem[{{Whitney} {et~al.}(2003){Whitney}, {Wood}, {Bjorkman}, \&
  {Wolff}}]{whitney2003}
{Whitney}, B.~A., {Wood}, K., {Bjorkman}, J.~E., \& {Wolff}, M.~J. 2003, \apj,
  591, 1049

\bibitem[{{Wolf} {et~al.}(2008){Wolf}, {Schegerer}, {Beuther}, {Padgett}, \&
  {Stapelfeldt}}]{wolf2008}
{Wolf}, S., {Schegerer}, A., {Beuther}, H., {Padgett}, D.~L., \& {Stapelfeldt},
  K.~R. 2008, \apjl, 674, L101

\bibitem[{{Woody} {et~al.}(2004){Woody}, {Beasley}, {Bolatto}, {Carlstrom},
  {Harris}, {Hawkins}, {Lamb}, {Looney}, {Mundy}, {Plambeck}, {Scott}, \&
  {Wright}}]{woody2004}
{Woody}, D.~P., {Beasley}, A.~J., {Bolatto}, A.~D., {Carlstrom}, J.~E.,
  {Harris}, A., {Hawkins}, D.~W., {Lamb}, J., {Looney}, L., {Mundy}, L.~G.,
  {Plambeck}, R.~L., {Scott}, S., \& {Wright}, M. 2004, in Society of
  Photo-Optical Instrumentation Engineers (SPIE) Conference Series, Vol. 5498,
  Society of Photo-Optical Instrumentation Engineers (SPIE) Conference Series,
  ed. C.~M. {Bradford}, P.~A.~R. {Ade}, J.~E. {Aguirre}, J.~J. {Bock},
  M.~{Dragovan}, L.~{Duband}, L.~{Earle}, J.~{Glenn}, H.~{Matsuhara}, B.~J.
  {Naylor}, H.~T. {Nguyen}, M.~{Yun}, \& J.~{Zmuidzinas}, 30--41

\bibitem[{{Yorke} \& {Bodenheimer}(1999)}]{yorke1999}
{Yorke}, H.~W., \& {Bodenheimer}, P. 1999, \apj, 525, 330

\end{thebibliography}
\end{small}

\clearpage

\clearpage

\clearpage

\clearpage

\clearpage

\clearpage

\clearpage

\clearpage

\clearpage

\clearpage

\begin{deluxetable}{llllll}
\tablewidth{0pt}
\rotate
\tabletypesize{\scriptsize}
\tablecaption{CARMA 3.4 mm Observations}
\tablehead{
  \colhead{Source} & \colhead{RA} & \colhead{Dec}  &\colhead{Config.}  & \colhead{Date} & \colhead{Calibrators}\\
             & \colhead{(J2000)} &  \colhead{(J2000)}    &  &    \colhead{(UT)}          &   \colhead{(Gain, Flux)}   \\
}
\startdata
L1527             & 04:39:53.9  & +26:03:09.6  & E-array & 02, 05, 06 Oct 2008 & 3C111, 3C84      \\
L1527             & ...  & ...  & D-array & 28, 30 Jul 2009 & 3C111, Mars     \\
L1527             & ...  & ...  & D-array & 31 Aug 2008 & 3C111, Uranus   \\
L1527             & ...  & ...  & C-array & 28 May 2009 & 3C111, 3C84    \\
L1527             & ...  & ...  & C-array & 11 Oct 2009 & 3C111, Uranus   \\
L1527             & ...  & ...  & B-array & 02 Jan 2010 & 3C111, Uranus      \\
L1527             & ...  & ...  & B-array & 06 Jan 2010 & 3C111, None      \\
L1527             & ...  & ...  & A-array & 02 Dec 2010 & 3C111, Neptune      \\
\enddata
\tablecomments{} 
\end{deluxetable}

\begin{deluxetable}{llllll}
\tablewidth{0pt}
\rotate
\tabletypesize{\scriptsize}
\tablecaption{SMA 870 \micron\ Observations}
\tablehead{
  \colhead{Source} & \colhead{RA} & \colhead{Dec}  &\colhead{Config.} & \colhead{Date} & \colhead{Calibrators}\\
             & \colhead{(J2000)} &  \colhead{(J2000)}    & &    \colhead{(UT)}          &   \colhead{(Gain, Flux)}   \\
}
\startdata
L1527             & 04:39:53.9  & +26:03:09.6  & Compact  & 17 Dec 2004 & 3C111, Uranus      \\
L1527             & ...  & ...  & Very Extended  & 05/06 Jan 2011 & 3C111, Callisto        \\

\enddata
\end{deluxetable}

\begin{deluxetable}{lllllllllll}

\tablewidth{0pt}
\tabletypesize{\scriptsize}
\tablecaption{L1527 Millimeter Fluxes}
\tablehead{
  \colhead{Wavelength} & \colhead{Configuration(s)} & \colhead{Integrated Flux} & \colhead{Peak Flux} & \colhead{Gaussian Size} & \colhead{Gaussian PA} & \colhead{Deconvolved Size} & \colhead{Deconvolved PA} & \colhead{Robust}& \colhead{Taper} & \colhead{Beam}\\
  \colhead{(mm)}        &                           & \colhead{(mJy)}           & \colhead{(mJy)}    & \colhead{(\arcsec)}    & \colhead{(\degr)}& \colhead{(\arcsec)}    & \colhead{(\degr)}  & &\colhead{(\arcsec)}  &\colhead{(\arcsec)} \\
}
\startdata
3.4 & A & 16.9$\pm$1.4 & 7.1$\pm$0.3 & 0.6 $\times$ 0.50 & -1.8 & 0.55 $\times$ 0.246 & 6.6 & 2 & ... & 0.43 $\times$ 0.33\\
0.87 & VEX & 213.6$\pm$8.1 & 80.8$\pm$5.7 & 0.55 $\times$ 0.24 & -1.5 & 0.54 $\times$ 0.14 & -3.6 & 2 & ... & 0.30 $\times$ 0.24\\
\\
3.4 & ABCDE & 22.5$\pm$1.8 & 20.5$\pm$0.6 & ... & ... & ... & ... & 2 & 2.0 & 2.2 $\times$ 2.0\\
0.87& VEX + COMP &235.5$\pm$5.8 & 160.6$\pm$4.8& ...& ...& ... & ... & 2 & ... & 0.62 $\times$ 0.49\\
\enddata
\tablecomments{Note that the uncertainties are only statistical.}
\end{deluxetable}

\begin{deluxetable}{llccll}
\tabletypesize{\scriptsize}
\tablewidth{0pt}
\tablecaption{Model Grid Parameters\label{grid}}
\tablehead{   \colhead{Parameter} & \colhead{Description} & \colhead{Values}}
\startdata
R$_{*}$(R$_{\sun}$) & Stellar radius & 2.09 \\
T$_{*}$(K) & Stellar temperature & 4000 \\
L$_{*}$(L$_{\sun}$) & System luminosity & 2.75\\
M$_{*}$(M$_{\sun}$) & Stellar mass & 0.5 \\
M$_{disk}$(M$_{\sun}$) & Disk mass & 0.005, 0.0075, 0.01, 0.025, 0.05, 0.075, 0.1    \\
R$_{disk,max}$/R$_{c}$(AU) & Disk Radius & 50.0, 100.0, 125.0, 150.0, 175.0, 200.0, 250.0, 300.0\\
H$_0$ (R$_{*}$) & Disk scale height at R$_{*}$ & 0.02, 0.025, 0.03, 0.035  \\
$p$ & Disk radial density exponent & 2.5, 3.0, 3.5, 4.0\\
$\gamma$ & Disk scale height exponent & 1.22, 1.25, 1.27, 1.3\\
$\dot{M}_{disk}$(M$_{\sun}$ $yr^{-1}$) &  Disk accretion rate & 3.0 $\times 10^{-7}$\\
R$_{trunc}$(R$_{*}$) & Magnetosphere co-rotation radius & 3.0 \\
F$_{spot}$ & Fractional area of accretion hotspot & 0.01\\
R$_{disk,min}$(R$_{*}$) & Disk inner radius & 14.25 \\
R$_{env,min}$(R$_{*}$) & Envelope inner radius & 42.75\\
R$_{env,max}$(AU) & Envelope outer radius & 15000\\
$\dot{M}_{env}$(M$_{\sun}$ $yr^{-1}$) & Envelope mass infall rate &  1.0 $\times 10^{-5}$ \\
b$_{out}$ & Outer cavity shape exponent & 1.5 \\
$\theta_{open,out}$($^{\circ}$) & Outer cavity opening angle & 20 \\
$\theta_{inc}$($^{\circ}$) & Inclination angle & 85\\
$\rho_{c}$(g cm$^{-3}$) & Cavity density & 0 \\
$\kappa_{850\mu m}$(cm$^{2}$ g$^{-1}$) & Dust opacity at 850\micron & 3.5 \\
$\beta_{dust}$ & Dust Spectral Index Disk/Envelope & 0.0 0.25 0.5 0.75\\

\enddata
\end{deluxetable}

\begin{deluxetable}{lllll}
\tabletypesize{\scriptsize}
\tablewidth{0pt}
\tablecaption{Model parameters\label{param}}
\tablehead{   \colhead{Parameter} & \colhead{Description} & \colhead{Paper I Model} & \colhead{Best Fit Model}  & \colhead{Parameter Use}}
\startdata
R$_{*}$(R$_{\sun}$) & Stellar radius    & 2.09 & 2.09  & fixed\\
T$_{*}$(K) & Stellar temperature        & 4000 & 4000  & fixed\\
L$_{*}$(L$_{\sun}$) & System luminosity & 2.75 & 2.75  & fixed\\
M$_{*}$(M$_{\sun}$) & Stellar mass      & 0.5  & 0.19\tablenotemark{\dagger} & fixed\\
M$_{disk}$(M$_{\sun}$) & Disk mass      & 0.005 & 0.0075 & varied \\
h(100AU) & Disk scale height at 100AU    & 36.3 & 48.0 & varied \\
H$_0$ & Disk scale height at R$_{*}$    & 0.03  & 0.03 & varied \\
$p$ & Disk radial density exponent      & 3.0   & 2.5  & varied \\
$\gamma$ & Disk scale height exponent    & 1.27 & 1.3 & varied \\
$\dot{M}_{disk}$(M$_{\sun}$ $yr^{-1}$)  &  Disk accretion rate & 3.0 $\times 10^{-7}$ & 1.5 $\times 10^{-6}$ &  fixed \\
R$_{trunc}$(R$_{*}$) & Magnetosphere co-rotation radius      & 3.0   & 3.0    & fixed\\
F$_{spot}$ & Fractional area of accretion hotspot            & 0.01  & 0.01   & fixed\\
R$_{disk,min}$(R$_{*}$) & Disk inner radius                  & 14.25 & 14.25  & fixed\\
R$_{disk,max}$/R$_{c}$(AU) & Disk outer radius               & 190   & 125    & varied \\
R$_{env,min}$(R$_{*}$) & Envelope inner radius               & 42.75 & 42.75 & fixed\\
R$_{env,max}$(AU) & Envelope outer radius                    & 15000 & 15000 & fixed\\
$\dot{M}_{env}$(M$_{\sun}$ $yr^{-1}$) & Envelope mass infall rate    & 0.8 $\times 10^{-5}$ & 4.5 $\times 10^{-6}$\tablenotemark{\dagger} & fixed \\
$\rho_{1AU}$(g cm$^{-3}$) & Envelope density at 1AU          & 5.8 $\times 10^{-14}$ & 7.25 $\times 10^{-14}$ & varied \\
b$_{out}$ & Outer cavity shape exponent                      & 1.5   & 1.5  & fixed \\
$\theta_{open,out}$($^{\circ}$) & Outer cavity opening angle & 20    & 20   & fixed \\
$\theta_{inc}$($^{\circ}$) & Inclination angle               & 85    & 85   & fixed\\
$\rho_{c}$(g cm$^{-3}$) & Cavity density                     & 0     & 0    & fixed\\
$\beta_{dust,mm}$ & Millimeter Dust Spectral Index           & ...   & 0.25 & varied\\
\enddata
\tablenotetext{\dagger}{The models were run with $M_*$ = 0.5 $M_{\sun}$; however, this value only goes
into the calculation of density from the infall rate and disk-protostar accretion rate. Since there is no real effect of the
reduced protostar mass on the model results, we simply rescale the accretion rate and infall rate to 
reflect the lower central mass. However, with the lower central mass, the protostar should be less luminous and the disk would need to be 
accreting more rapidly. Given that the sub/millimeter emission is all reprocessed, the source of luminosity being 
accretion or photospheric is not an important distinction.}
\end{deluxetable}

\begin{deluxetable}{llllllll}
\tabletypesize{\scriptsize}
\tablewidth{0pt}
\tablecaption{Parameter Ranges From Fitting}
\tablehead{\colhead{Weighted Data} & \colhead{Number of Models} & \colhead{R$_{disk}$} & \colhead{M$_{disk}$} & \colhead{$\alpha_{disk}$} & \colhead{$H(r)$} & \colhead{$H_{init}$} & \colhead{$\beta$}\\
   &  & \colhead{(AU)} & \colhead{($M_{\sun}$)}&  & & \colhead{($R_{*}$)}& 
}
\startdata
Uniform Weighting & 032 & 125$^{200}_{100}$ & 0.0075$^{0.010}_{0.005}$ & 2.5$^{3.0}_{2.5}$ & 1.30$^{1.30}_{1.25}$ & 0.030$^{0.035}_{0.020}$ & 0.25$^{0.25}_{0.00}$\\\\
SED, IRS, Visibilities, and Sub/mm images & 052 & 100$^{300}_{100}$ & 0.0100$^{0.100}_{0.007}$ & 2.5$^{3.5}_{2.5}$ & 1.25$^{1.30}_{1.22}$ & 0.025$^{0.035}_{0.020}$ & 0.25$^{0.50}_{0.25}$\\\\
Visibilities and Sub/mm images & 056 & 100$^{300}_{100}$ & 0.0075$^{0.100}_{0.005}$ & 2.5$^{3.5}_{2.5}$ & 1.27$^{1.30}_{1.22}$ & 0.035$^{0.035}_{0.020}$ & 0.25$^{0.50}_{0.00}$\\\\
Visibilities & 120 & 100$^{300}_{100}$ & 0.0075$^{0.100}_{0.005}$ & 2.5$^{3.5}_{2.5}$ & 1.27$^{1.30}_{1.22}$ & 0.030$^{0.035}_{0.020}$ & 0.25$^{0.75}_{0.25}$\\\\
Sub/mm images & 024 & 100$^{175}_{100}$ & 0.0075$^{0.025}_{0.005}$ & 2.5$^{3.0}_{2.5}$ & 1.27$^{1.30}_{1.22}$ & 0.035$^{0.035}_{0.020}$ & 0.25$^{0.50}_{0.00}$\\\\
SED and IRS spectrum & 209 & 200$^{300}_{100}$ & 0.0100$^{0.075}_{0.005}$ & 3.0$^{3.5}_{2.5}$ & 1.27$^{1.30}_{1.22}$ & 0.025$^{0.035}_{0.020}$ & 0.00$^{0.75}_{0.00}$\\\\
SED, IRS spectrum, and L\arcmin\ image & 037 & 150$^{175}_{125}$ & 0.0075$^{0.010}_{0.005}$ & 3.0$^{3.0}_{2.5}$ & 1.27$^{1.30}_{1.25}$ & 0.030$^{0.035}_{0.025}$ & 0.00$^{0.50}_{0.00}$\\\\
L\arcmin\ image & 016 & 125$^{300}_{125}$ & 0.0050$^{0.100}_{0.005}$ & 3.0$^{4.0}_{2.5}$ & 1.30$^{1.30}_{1.27}$ & 0.035$^{0.035}_{0.025}$ & 0.00$^{0.25}_{0.00}$\\\\
L\arcmin\ and Sub/mm images & 045 & 150$^{200}_{100}$ & 0.0075$^{0.010}_{0.005}$ & 2.5$^{3.0}_{2.5}$ & 1.27$^{1.30}_{1.25}$ & 0.030$^{0.035}_{0.020}$ & 0.00$^{0.25}_{0.00}$
\enddata

\tablecomments{We show the parameters of the models with $\chi^2$ - $\chi_{best}^2$ $<$ 3 \citep{rob2007};
 except for the L\arcmin\ image and L\arcmin\ + sub/mm images, where we chose 
$\chi^2$ - $\chi_{best}^2$ $<$ 8 in order to get a larger sampling of models. To 
give certain datasets more weight, we multiplied the respective $\chi^2$ by 10 The best fitting 
value is given in each column with the upper and lower ranges as superscripts and subscripts. Looking at the fits
in this way better shows which sets of data constrain which values.  It is clear that the data
which do not take into account the spatial structure of the source (visibilities, SED, and IRS spectrum) do not
provide adequate constraints on the models. The images do provide the best constraints on the source properties, but the
additional constraints from the visibilities and SED/IRS are able to rule out some additional models. 
}
\end{deluxetable}

\begin{deluxetable}{lllll}
\tabletypesize{\scriptsize}
\tablewidth{0pt}
\tablecaption{Model Parameter Comparisons}
\tablehead{   \colhead{Parameter} &\colhead{Description} & \colhead{L1527} & \colhead{CB26} & \colhead{IRAS 04302+2247}  }
\startdata
M$_{disk}$(M$_{\sun}$) & Disk mass         & 0.0075 & 0.3 & 0.07 \\
h(100AU) & Disk scale height at 100 AU     & 48     & 10    & 15   \\
$p$ & Disk radial density exponent         & 2.5    & 2.2   & 2.37 \\
$\gamma$ & Disk scale height exponent      & 1.3    & 1.27  & 1.29 \\
R$_{disk,max}$(AU) & Disk outer radius     & 125    & 200   & 300   \\
$\kappa_{dust,850\mu m}$ (cm$^2$ g$^{-1}$)& Millimeter Dust Opacity  & 3.5   & 2.5  & 2.5     \\
$\beta_{dust,mm}$ & Millimeter Dust Spectral Index  & 0.25   & ...  & ...     \\
\enddata
\end{deluxetable}

\end{document}